\documentclass[useAMS,usenatbib]{mn2e}
\usepackage{times}
\usepackage{url}
\usepackage{graphics,graphicx}
\usepackage{deluxetable}
\usepackage[usenames]{color}
\DeclareGraphicsExtensions{.pdf,.png,.jpg,.mps,.eps,.ps}
\usepackage{amsmath}
\usepackage{lscape}
\usepackage{natbib}
\usepackage[dvipsnames]{xcolor}
\usepackage{tikz}

\newcommand\astrosat{{\it ASTROSAT}}
\newcommand\suzaku{{\it Suzaku}}
\newcommand\asca{{\it ASCA}}
\newcommand\sax{{\it BeppoSAX}}

\newcommand\rosat{{\it ROSAT}}

\newcommand\xmm{{\it XMM--Newton}}
\newcommand\swift{{\it Swift}}

\newcommand\ks{{\rm~ks}}

\newcommand\kev{{\rm~keV}}
\newcommand\ev{{\rm~eV}}
\newcommand{\xiunit}{$\rm erg~cm~s^{-1}$}
\newcommand{\logxi} {$\rm {log}(\xi/erg~cm~s^{-1}$)}

\newcommand\kms{\ifmmode {\rm~km\ s}$^{-1}$ \else ~km s$^{-1}$\fi}
\newcommand\Hunit{\ifmmode {\rm~km\ s}$^{-1}$\ {\rm Mpc}$^{-1}$
        \else ~km s$^{-1}$ Mpc$^{-1}$\fi}
\newcommand\ctssec{\ifmmode {\rm~count\ s}$^{-1}$ \else ~count s$^{-1}$\fi}
\newcommand\ergsec{\ifmmode {\rm~erg\ s}$^{-1}$ \else
        ~erg s$^{-1}$\fi}
\newcommand\funit{\ifmmode {\rm~erg\ s}$^{-1}$\ ; {\rm cm}$^{-2}$ \else
        ~ergs s$^{-1}$ cm$^{-2}$\fi}
\newcommand\phflux{\ifmmode {\rm~photon\ s}$^{-1}$\  ; {\rm cm}$^{-2}$
        \else   ~photon s$^{-1}$ cm$^{-2}$\fi}
\newcommand\efluxA{\ifmmode {\rm~erg\ s}$^{-1}$\ ; {\rm cm}$^{-2}$\ ; {\rm
        \AA}$^{-1}$ \else ~erg s$^{-1}$ cm$^{-2}$ \AA$^{-1}$\fi}
\newcommand\efluxHz{\ifmmode {\rm~erg\ s}$^{-1}$\ ; {\rm cm}$^{-2}$\ ; {\rm
        Hz}$^{-1}$ \else ~erg s$^{-1}$ cm$^{-2}$ Hz$^{-1}$\fi}
\newcommand\cc{\ifmmode {\rm~cm}$^{-3}$ \else cm$^{-3}$\fi}
\newcommand\FWHM{\ifmmode {\rm~FWHM} \else ${\rm~FWHM}$\fi}
\newcommand\Msun{\ifmmode M_{\odot} \else $M_{\odot}$\fi}
\newcommand\Lsun{\ifmmode L_{\odot} \else $L_{\odot}$\fi}

\newcommand\hbeta{\ifmmode {\rm H}\beta \else H$\beta$\fi}
\newcommand\Kalpha{\ifmmode {\rm K}\alpha \else K$\alpha$\fi}
\newcommand\nh{\ifmmode N_{\rm H} \else N$_{\rm H}$\fi}

\makeatletter

\newcommand{\Rmnum}[1]{\expandafter\@slowromancap\romannumeral #1@}
\makeatother

\title { Complex Optical/UV and X-ray Variability of the Seyfert 1 galaxy 1H~0419--577}
\author [Pal et al.] {Main Pal$^{1,2}$\thanks { Email: mainpal@prl.res.in}, Gulab C. Dewangan$^2$\thanks{Email: gulabd@iucaa.in}, 
Ajit K. Kembhavi $^2$, Ranjeev Misra$^2$\thanks{Email: rmisra@iucaa.in}
\newauthor Sachindra Naik $^{1}$\\	\\
$^{1}$Astronomy \& Astrophysics Division, Physical Research Laboratory, Ahmedabad - 380009, India.\\
$^{2}$Inter University Centre for Astronomy and Astrophysics, Pune - 411007, India.
}

\begin{document}
\maketitle
\begin{abstract}
	We present detailed broadband UV/optical to X-ray spectral variability of the Seyfert 1 galaxy 1H~0419--577 using six \xmm{}~observations performed during 2002--2003. These observations covered a large amplitude variability event in which the soft X-ray (0.3-2 keV) count rate increased by a factor of $\sim 4$ in six months. The X-ray spectra during the variability are well described by a model consisting of a primary power law, blurred and distant reflection. The 2-10 keV power-law flux varied by a factor $\sim$7 while the 0.3-2 keV soft X-ray excess flux derived from the blurred reflection component varied only by a factor of $\sim2$. The variability event was also observed in the optical and UV bands but the variability amplitudes were only at the 6-10\% level. The variations in the optical and UV bands appear to follow the variations in the X-ray band. During the rising phase, the optical bands appear to lag behind the UV band but during the declining phase, the optical bands appear to lead the UV band. Such behavior is not expected in the reprocessing models where the optical/UV emission is the result of reprocessing of X-ray emission in the accretion disc. The delayed contribution of the broad emission lines in the UV band or the changes in the accretion disc/corona geometry combined with X-ray reprocessing may give rise to the observed behavior of the variations.

 \end{abstract}

\section{Introduction}
Active galactic nuclei (AGN) exhibit variable and complex spectral energy distribution (SED).  The central engine of an AGN is known to radiate in the luminosity range $10^{41-47}~\rm ergs~s^{-1}$ almost equally over the entire electromagnetic spectrum. In the brightest AGN, the intense radiation emitted from the central engine dominates the light coming from other constituents of the host galaxy such as stars. The huge energy release is attributed to accretion onto a supermassive black hole (SMBH) \citep{1963MNRAS.125..169H, 1969Natur.223..690L}. The emitting regions in the vicinity of an SMBH are too small to be identified separately by any available instruments on ground or space facilities. Therefore, the multiwavelength variability of spectral components such as  the big blue bump, power--law continuum, reflection components (i.e., Fe-K$\alpha$~and reflection hump), soft X-ray excess and  relationship between them are the most effective way to investigate the complex phenomena in the exotic physical conditions in the nuclear environment of an AGN. Indeed, variability studies of AGN over the last few decades have been important to understand the physical structure of emitting regions close to the SMBH.

The most pronounced SED component in the UV/optical band of Seyfert 1 galaxies is the ``Big Blue Bump" (BBB). The BBB is thought to
arise from the thermal emission from standard accretion disc peaking in the UV regime
\citep[see e.g.,][]{1973A&A....24..337S, 1978Natur.272..706S, 1999PASP..111....1K}. According to the standard accretion disc theory, the outer
disc emits in the optical band while the inner disc emits at lower
wavelengths in the UV, with the emission sometimes extending to the softest X-rays \citep{2013MNRAS.434.1955D,
2015MNRAS.446..759C}. In such a scenario, one would expect to observe emission
from different parts of the accretion discs with a wavelength dependent time lag. One should be able to detect lag or lead between the optical/UV to X-ray emission arising from different regions. The optical/UV emission arising from the accretion disc can lag behind or lead the X-ray emission from a compact hot corona depending on whether X-ray reprocessing or intrinsic disc emission dominates the optical/UV band. The origin of UV/optical emission variability has been a subject of intense research.  Recently \citet{2013MNRAS.433.1709G} have found that the optical emission leads UV emission and UV emission leads soft X-rays in a radio loud
narrow line Seyfert 1 galaxy PKS~0558--504, supporting a propagation fluctuation model.
In reprocessing scenario, one would expect that the UV/optical bands lag the X-
rays (e.g., NGC5548: \citealt{2014MNRAS.444.1469M, 2015ApJ...806..129E}).

Another ubiquitous feature of the SED of an AGN in X-ray is power-law continuum emission. 
This emission is believed to be non-thermal emission from the central engine originating 
through inverse Compton upscattering of soft photons from the accretion disc in the hot 
corona (\citealt{1980A&A....86..121S,1991ApJ...380L..51H}). Often power-law continuum 
arises with a Fe--K$\alpha$ reflection component near 6 keV and a reflection hump in the 
$10-40\kev$ band. The Fe-K$\alpha$~emission line arises due to  photoelectric absorption followed by fluorescence emission.
The reflection hump peaking near 20 keV is the result of the photoelectric absorption at low energies  and Compton back scattering
in an optically thick material such as a putative torus or an accretion disc \citep{1988MNRAS.233..475G,
1988ApJ...335...57L, 1991MNRAS.249..352G}. The low energy band below about 2 keV  in X-rays is sometimes dominated by a smooth component, known as the  ``soft X-ray excess (SE) emission'' over the broadband power law component. This SE component was discovered in Seyfert type 1 galaxies in 1980s \citep{1985ApJ...297..633S,1985MNRAS.217..105A}. The origin of this component is still debated, but is most likely the blurred reflection from a partially ionized accretion disc \citep{2005MNRAS.358..211R,2006MNRAS.365.1067C, 2013MNRAS.428.2901W}) or  thermally Comptonized emission from an optically thick, warm ($kT_e \sim 0.3\kev$) medium which is possibly the inner accretion disc (e.g., \citealt{1998MNRAS.301..179M, 2009MNRAS.398L..16J}). These models are found to describe the observed soft excess equally well (e.g., \citealt{2005AIPC..774..317S, 2007ApJ...671.1284D}).

Information on time delay between power--law emission and soft band emission (i.e., soft X-ray and UV/optical) plays a crucial role in determining the driving mechanism. Sometimes soft emissions are observed before the hard emission such as power--law emission. This is termed as `hard lag' \citep[see e.g.,][]{2001ApJ...554L.133P, 2004MNRAS.348..783M,2007ApJ...671.1284D,2013ApJ...777L..23W} and it is not well
understood. It can be described by a propagation fluctuation model in
which the variations in accretion flow are modulated towards the inner
accretion disc on the viscous timescales and finally the variations appear in strongly variable coronal X-ray emission \citep{1997MNRAS.292..679L,
2001MNRAS.327..799K}. More recently, a new type of lag called soft lag, where soft photons are received after hard photons, has been found in a number of AGN \citep{2009Natur.459..540F,2010MNRAS.401.2419Z,2013ApJ...764L...9C,
2013MNRAS.429.2917F, 2013MNRAS.431.2441D}. The UV/optical emission from the accretion disc have been reported to lag the X-ray emission via X-ray reprocessing ( see e.g., \citealt{2014MNRAS.444.1469M,2016ApJ...821...56F, 2017MNRAS.464.3194B, 2012MNRAS.422..902C, 2010MNRAS.403..605B, 2009MNRAS.394..427B}). Optical/UV lags resulting from X-ray reprocessing can be used to probe the nature of accretion discs as suggested by \citet{2014MNRAS.444.1469M}. Optical/UV leads, on the other hand, provide important information to probe the propagation fluctuation model for the energy dependent variability (e.g., \citealt{2003MNRAS.343.1341S, 2008ApJ...677..880M}). Variations in the optical/UV and X-ray bands can also probe the variable partial covering absorption sometimes invoked to explain the observed spectral variability from AGN. Thus, simultaneous observations in X-ray and UV/optical bands can provide important clues on the complex environment around the central engine. 1H~0419-577 is a unique AGN to investigate the complex mechanisms for the variability.

1H~0419--577 is a well known broad-line Seyfert 1 AGN located at a redshift
$z=0.104$ \citep{briss1987, grupe96,guainazzi1998,turner1999}. It has been
observed by almost all X-ray satellites such as \rosat{}, \sax{}~and \asca{} 
in the past and in the present decade by \xmm{}, \suzaku{}~and \swift{}. This 
AGN has shown complex X-ray broadband spectra studied by various authors
\citep{2010MNRAS.408..601W, 2009ApJ...698...99T, fabian2005, pounds2004a, 
pounds2004b, 2002MNRAS.330L...1P}. In the X-ray spectrum, this AGN has 
revealed strong soft X-ray excess below 2\kev, moderate Fe-K$\alpha$ near 6 keV  and reflection hump
in the $\sim 10-40$ keV band \citep{2009ApJ...698...99T, 2010MNRAS.408..601W, 2013MNRAS.435.1287P}. 
\citet{2014A&A...563A..95D} presented the historical spectral variability based on broadband SED fitting using disc blackbody with a temperature $\sim56$ eV for the UV/Optical emission and two Comptonization components (warm plasma with $\tau$ $\sim$ 7, $kT_e$ $\sim$ 0.7 keV for soft excess and hot plasma with $\tau$ $\sim$ 0.5, $kT_e$ $\sim$ 160 keV) for X-ray emission modified by a partial covering neutral absorber. They claimed that the spectral variability may be explained by the variable partially covering neutral absorber having the column density in the range $N_{\rm H}\sim10^{19-22}~\rm cm^{-2}$.  Using two \suzaku{}~and two \xmm{} observations, \citet{2013MNRAS.435.1287P} showed that the full broadband complex spectra are best described by the blurred reflection model, which was supported by the observed spectral variability. The implied strong X-ray illumination should also give rise to reprocessed emission in the optical/UV bands. The UV/optical emission of this AGN are found variable (see e.g., \citealt{2014A&A...563A..95D, pounds2004b}) and their detailed relationship with the X-ray emission is explorable.

The nature of the variability of the optical/UV emission and its connection with the X-ray emission is not clearly understood. For example, NGC~3516 has shown fluke relationship between the X-ray and the UV/Optical radiation in various studies suggesting no clear evidence of X-ray connection with UV/optical emission \citep{2000AJ....119..119M, 2000ApJ...534..180E, 2002AJ....124.1988M,2015MNRAS.453.3455R, 2017MNRAS.464.3194B}. However, recent studies have shown the relationship between the optical/UV and the X-rays through either the X-ray reprocessing (e.g., NGC5548: \citealt{2014MNRAS.444.1469M, 2015ApJ...806..129E}) or likely propagation fluctuation model (e.g., PKS~0558--504 : \citealt{2013MNRAS.433.1709G}).
1H~0419--577 is well known to be highly variable in the X-rays (see e.g., \citealt{2013MNRAS.435.1287P}) as well as a UV bright source \citep{1995MNRAS.274.1165P,2013A&A...550A..71V}. The available multi-epoch data sets of this AGN are well suited for the study of the complex relationship between optical/UV and X-ray variability which is not well understood. Thus, available simultaneous data sets in X-ray to UV/optical bands and luminous nature in UV and X-rays make this AGN an ideal laboratory to investigate the origin of the UV/optical emission variability and its relationship with various X-ray spectral components. 

We organize the paper as follows. In the next section, we describe the
observations and data reduction. We present the data analysis in
Section 3 and discuss the main results in Section 4. We use the cosmological parameters $H_{0} = 67.04~{\rm
km~s^{-1}~Mpc^{-1}}$, $\Omega_m = 0.3183$ and $\Omega_{\Lambda} =
0.6817$ \footnote{http://www.kempner.net/cosmic.php} to calculate distance.

\section{Observation and data reduction}

 \xmm{} has observed 1H~0419--577 nine times since its launch. All available observations have been studied by several authors \citep[e.g.,][]{2014A&A...563A..95D, 2013MNRAS.435.1287P, 2013A&A...556A..94D, tombesi2012, 2010A&A...519A..17V, tombesi2010, 2010MNRAS.408..601W, 2009ApJ...698...99T, fabian2005, pounds2004a, pounds2004b, 2002MNRAS.330L...1P}. 
 Here we study the relationship between optical/UV and X-ray variability of 1H~0419--577 using six monitoring observations performed during 2002-2003 (see Table~\ref{obs_log}).

    We followed standard reduction procedure using the \xmm{} Science Analysis System ({\tt SAS version 14.0}: \citealt{2004ASPC..314..759G}) with the latest calibration files. We utilized all observations listed in Table~\ref{obs_log} with Optical Monitor \citep[{OM}:][]{2001A&A...365L..36M}, onboard \xmm{}~to study the UV/optical emission. 1H~0419--577 has been observed by the Optical Monitor with five optical/UV filters (V, B, U, UVW1, UVW2) in each of the observation during the 2002-2003  monitoring campaign.  The exposure time for each filter was in the range of 25 to 50 minutes. We reprocessed the OM data sets using {\tt SAS} tool {\tt omichain} and used all available UV/optical data sets. We obtained background corrected count rates from the source list for each OM filter and converted the count rates into respective flux densities~\footnote{http://xmm.esac.esa.int/external/xmm\_user\_support/documentation/sas\_usg/\\USG/ommag.html}.   

We also reprocessed the European Photon Imaging Camera (EPIC)-pn \citep{struder2001} data, contemporaneous with the OM observations, using {\tt epproc} and obtained event files. After examining the presence of flaring particle background by extracting light curves above $10\kev$, we removed the intervals which were affected by flares. The observations of orbit numbers 558 and 720 were found significantly affected by solar/proton flares. We, therefore, used higher count rate cutoffs to get sufficient counts for spectral modeling. In case of the observation of orbit 720, we obtained useful exposure time of only $\sim0.3\ks$ after excluding the intervals of high particle background. We used single and double events (PATTERN $\le4$) for the EPIC-pn, and omitted events in bad pixels and at the edges of the CCD (FLAG=0). We extracted the source spectrum from a circular region with radius 40 arcsec centred at the source. We obtained background spectra from the circular regions with radii in the $30-83$ arcsec range away from the source and without any bright portion of the CCD. The resultant net counts of each observation are listed in Table~\ref{obs_log}. We did not find any significant pileup which might affect our analysis. We generated response matrix and ancillary response files at the source position using the tools {\tt  rmfgen} and {\tt arfgen}, respectively and grouped the data using the SAS task {\tt specgroup} with an oversampling of $5$ and minimum counts of $20$ per bin.

\begin{table*}
  \centering
  \caption{Observations of 1H~0419--577 } \label{obs_log}
  \begin{tabular}{lcccccccc}
    \hline
    \hline
    Obs. ID          &  Date      & Orbit  & Optical/UV Filter$^{a}$    & On Target  & Net counts         & Rate$^{b}$              \\
                     &            &        &                       & Exposure (ks)         &             &${\rm counts~s^{-1}}$   \\

    \hline
     0148000201        & September 25, 2002 & 512 & V, B, U, W1, W2   & 15.1   &25347     & $2.27\pm0.01$           \\
     0148000301        & December 27, 2002  & 558 &  "                & 17.9  &13424      & $3.49\pm0.03 $            \\
     0148000401        & March 30, 2003     & 605 &  "                & 13.9  &86082      & $7.79\pm0.03 $             \\
     0148000501        & June 25, 2003      & 649 &  "                & 13.2  &67796      & $6.48\pm0.03 $              \\
     0148000601        & September 16, 2003 & 690 &  "                & 13.9  &65406      & $5.83\pm0.02 $              \\
     0148000701        & November 15, 2003  & 720 &  "                & 12.2  &2459       & $8.3\pm0.2 $                 \\
  \hline
  \end{tabular}\\
$^{a}$ W1= UVW1, W2= UVW2, $^b$  The EPIC-pn counts and count rates (counts s$^{-1}$) are quoted in the $0.3-10\kev$ band. 
\end{table*}

\begin{figure}\centering
\includegraphics[scale=0.6]{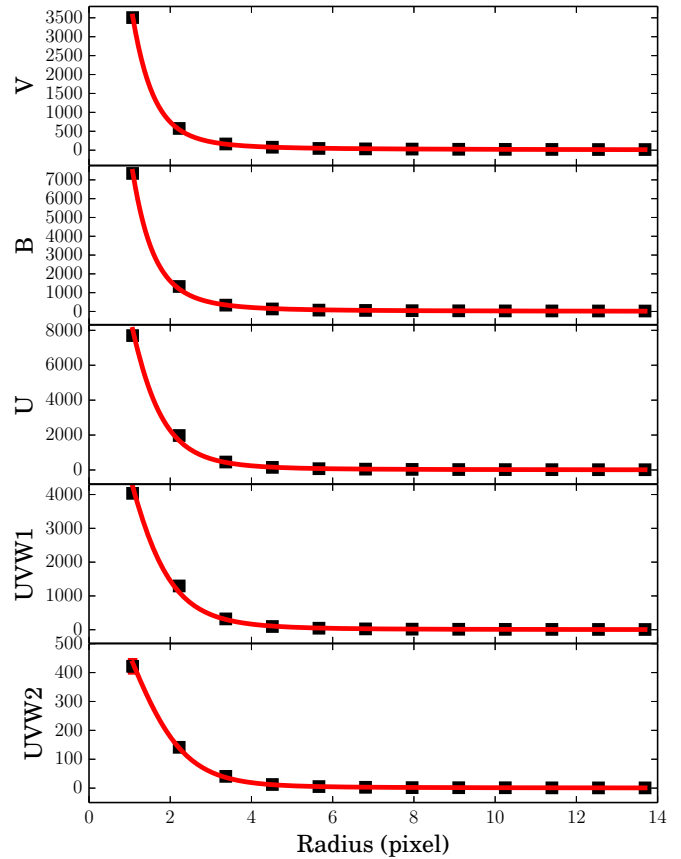}
\caption{Surface brightness ($\rm counts~pixel^{-2}$) values extracted from the observed PSF for each band are shown. The solid curves represent the best-fit model consisting of the Moffat function for AGN component and Sersic function for stellar constituents.} 
  \label{psfFit}	
\end{figure}

\begin{figure*}
  \centering
  \includegraphics[height=10cm,width=7.5cm]{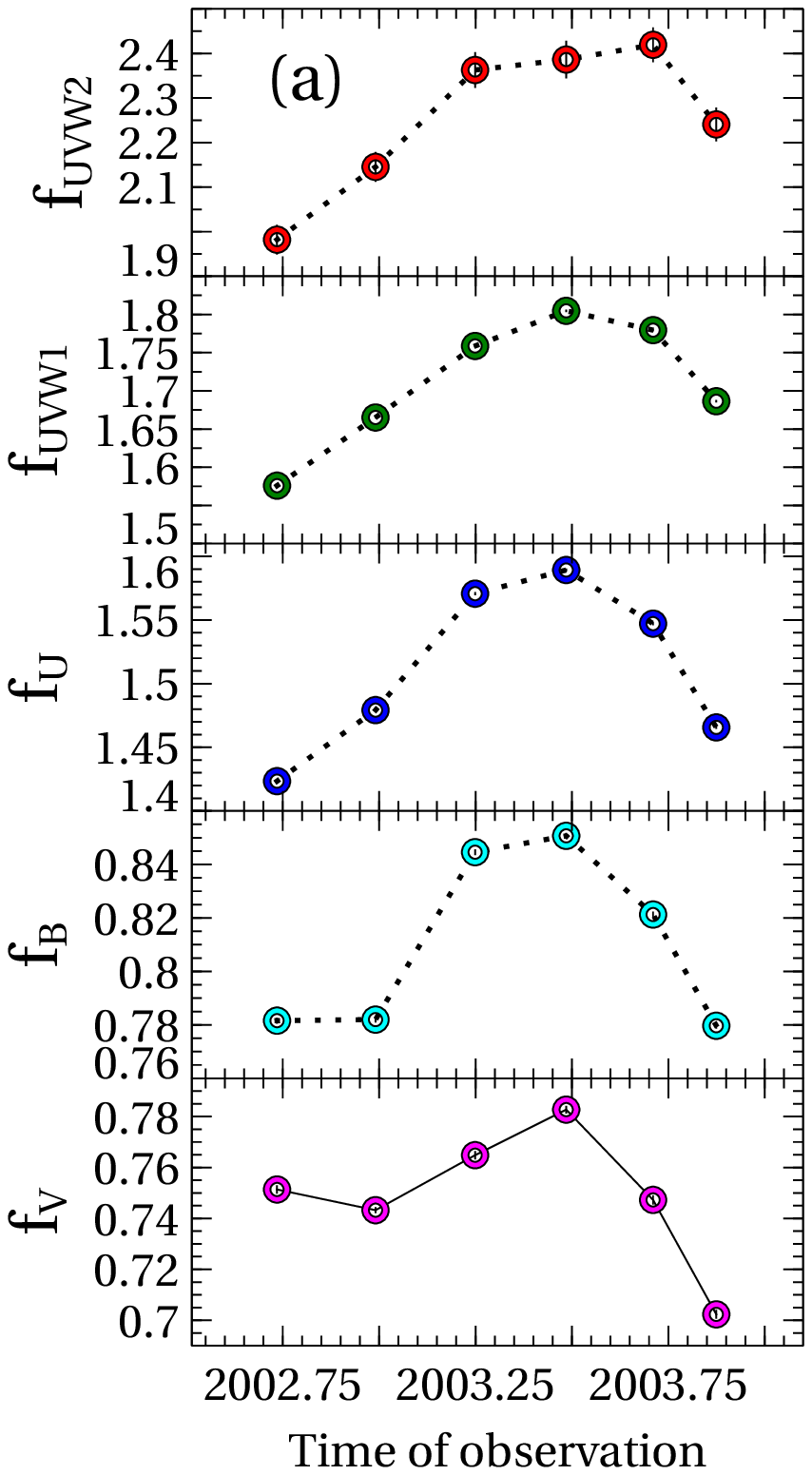} 
  \includegraphics[height=10cm,width=6.5cm]{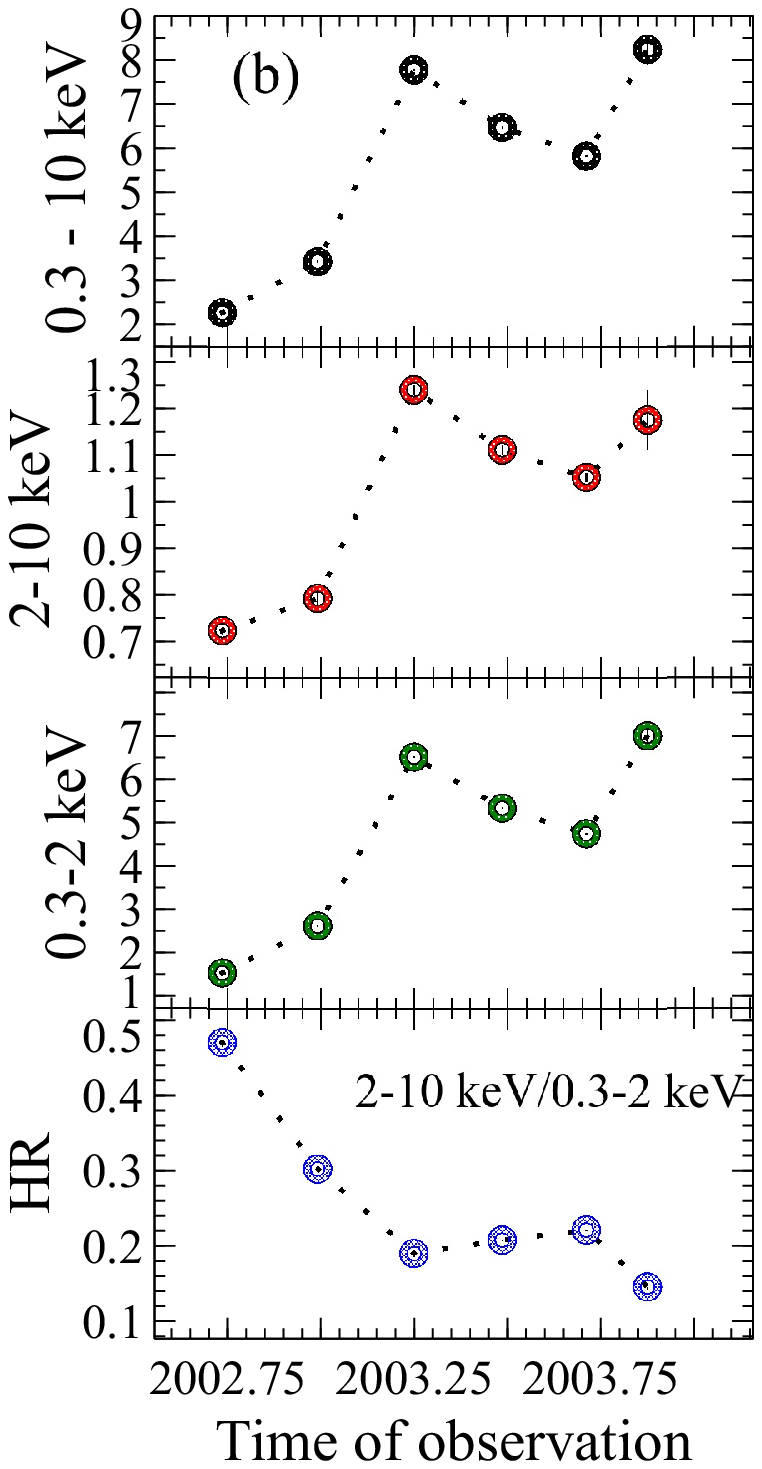}
  \caption{ (a) The light curves of the optical/UV flux density in various bandpass, where flux density is measured in units of $10^{-14}$\efluxA. (b) The light curves of X-ray count rates (counts~s$^{-1}$) in various energy bands and hardness ratio (HR) between $2-10$ keV and $0.3-2$ keV bands.}
  \label{uvopt}	
\end{figure*}

\begin{figure}\centering
  \includegraphics[scale=0.5]{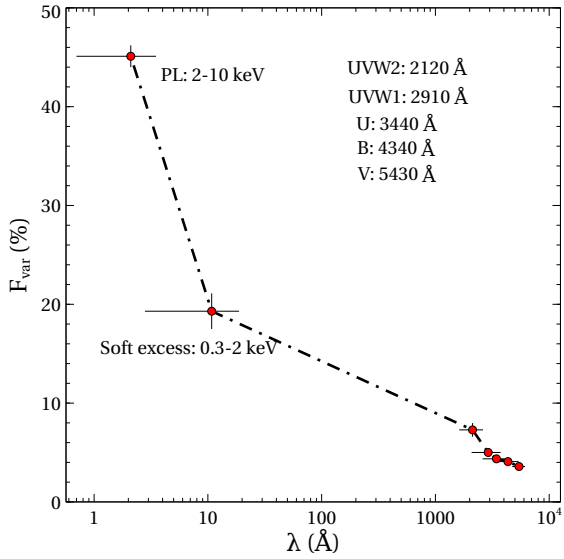}
\caption{ The fractional variability amplitude $F_{var}$ as a function of wavelength from X-ray to UV/optical bands. The $F_{var}$ decreases with increasing wavelength.} 
  \label{fvar}	
\end{figure}

\begin{figure}
\includegraphics[scale=0.43]{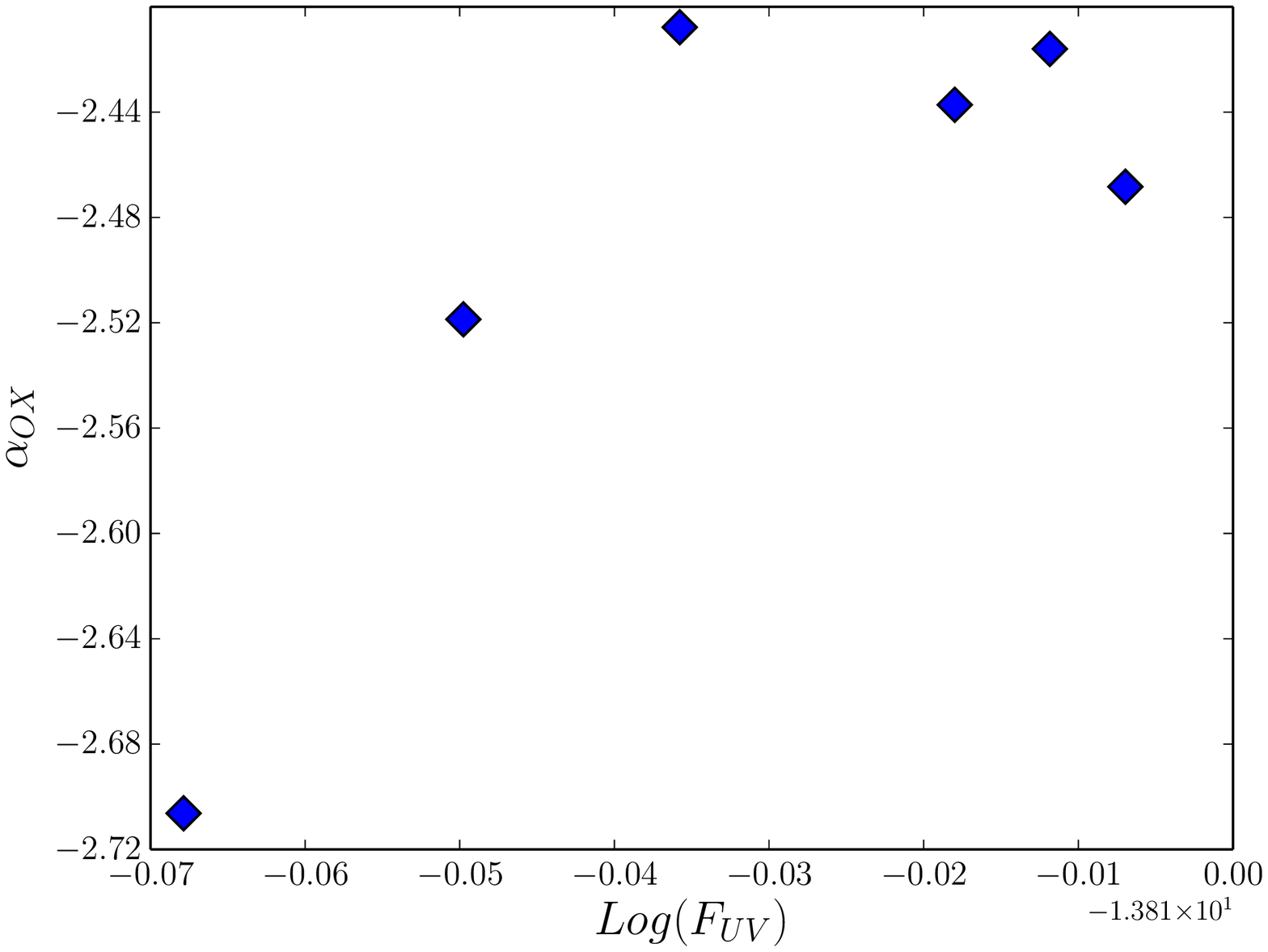}
\includegraphics[height=8.5cm,width=9.5cm]{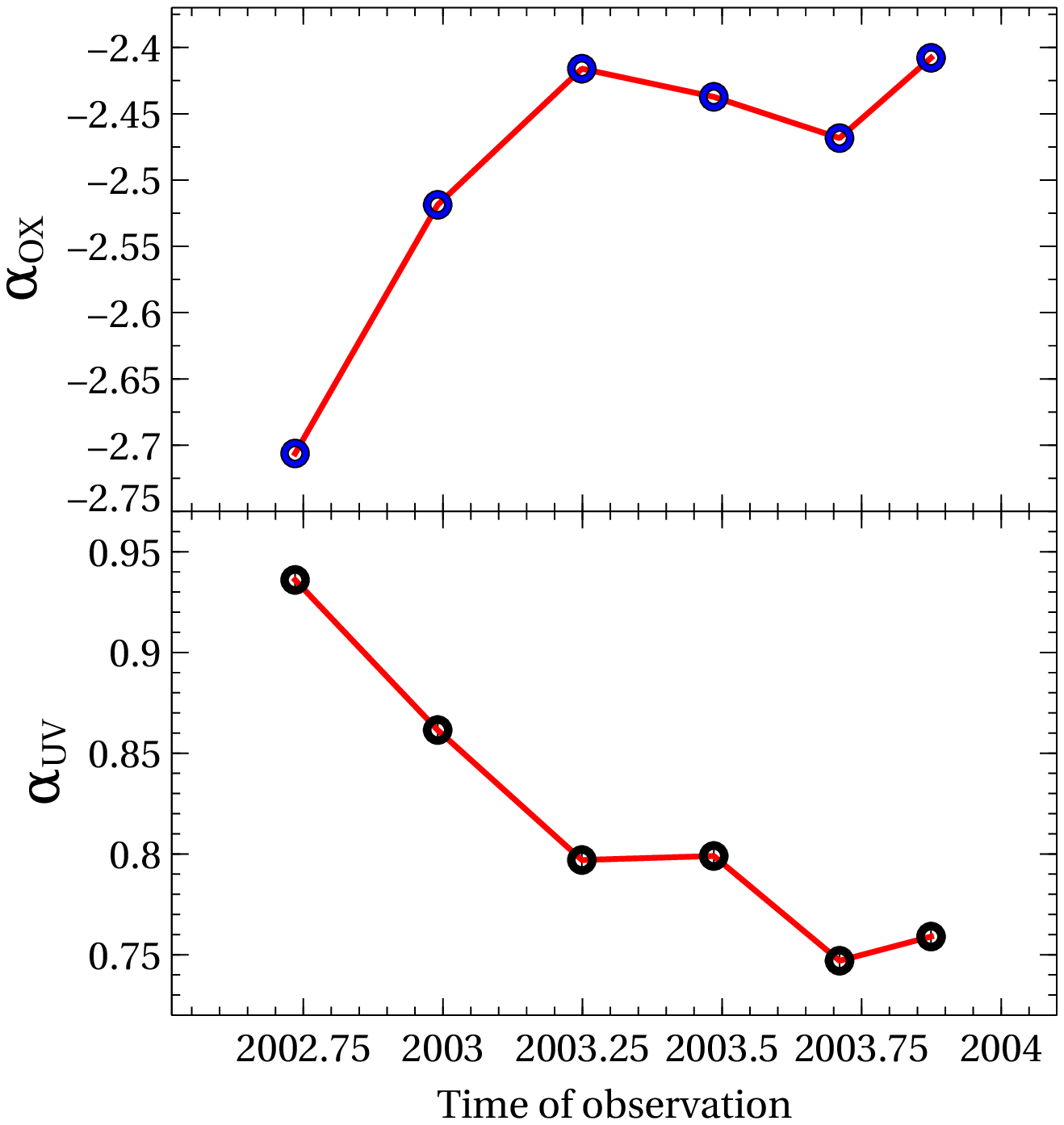}
\caption{{\it Top panel:} The $\alpha_{\rm{OX}}$ as a function of $log(F_{UV})$ at 2500\AA. {\it Bottom panel:} Variation of $\alpha_{\rm{OX}}$ (upper panel) and $\alpha_{\rm{UV}}$ (lower panel) during 2002-2003. 
Data points for $\alpha_{\rm OX}$ are shown without error bars (as mentioned in the text) whereas the values of $\alpha_{\rm{UV}}$ are shown with error bars. } 
  \label{oxuv}	
\end{figure}

\section{Data analysis}
To study the optical/UV emission variability and its connection with the X-ray emission, we used various techniques for timing as well as broadband X-ray spectroscopic analysis. Before carrying out our analysis, we corrected the UV/optical flux densities for contaminations from the host galaxy as described below.

\subsection{Decomposition of flux components}
 
The observed UV/optical flux can be considered as the combination of various emission components e.g., central engine, bulge and the disc of host galaxy, broad line regions (BLR) and narrow line regions (NLR). For this purpose, we used the observation of a white dwarf BPM~16274 (ID: 0701581601) performed with the Optical Monitor to derive the Moffat function parameters (see Equation ~\ref{moffat}). These parameters can be further used to estimate the flux from a point source such as 1H~0419$-$577. We modeled the radial profile (surface brightness as a function of detector pixel) of the observed point spread function (PSF) of the white dwarf by using Moffat function as given below 
\begin{equation}\label{moffat}
I(r)=I_{0}(1+(\frac{r}{a})^2)^{-b}
\end{equation}
 where $I$, $I_0$, $a$, $b$ are observed surface brightness at $r$, maximum brightness at $r=0$, scale factor and slope, respectively. We also used a constant for the background brightness present in the PSF of the white dwarf. After fitting Moffat function plus a constant to the radial profile of the PSF for each filter, the best-fit parameters ($a$, $b$) obtained are listed in Table~\ref{psfDecomp}.

We then fixed the best-fit values of $a$ and $b$ of the Moffat function and modeled the radial profile of the observed PSF from the OM observations (UVW2, UVW1, U, B and V) of 1H~0419$-$577. The background subtracted radial profiles were fitted with the Moffat function for the AGN contribution. We also included the Sersic function for stellar contribution \citep{2005PASA...22..118G} as described below 
\begin{equation}\label{sersicFunc}
I_s=I_{bd}~exp(-(1.9992n-0.3271)((r/r_{bd})^{1/n}-1))
\end{equation}
 where $I_s$ is the surface brightness at a distance $r$ from the centre, $I_{bd}$ is the maximum brightness due to the bulge and disc at a distance $r_{bd}$, $n$ is the Sersic index for galactic bulge and disc contribution. We obtained the best-fit parameters (I$_{bd}$, r$_{bd}$ and $n$) required to decompose the emission from the AGN and the bulge plus disc contribution. We then integrated both Moffat and Sersic functions from $r=0$ to $r=15$ pixels to estimate the net surface brightness for each component for every band. The contribution from AGN and the host galaxy are listed in Table~\ref{psfDecomp}. The host galaxy contribution is only at the level of $5-7\%$. The best-fit to the radial profiles of the PSF of AGN for each band are shown in Fig.~\ref{psfFit}. 

In order to remove the contribution of emission lines from BLR and NLR, we used the equivalent width of all possible emission lines listed in \citet{2001AJ....122..549V} in the observed bands. Using the typical band–width of OM filters, we estimated the net contribution of emission lines in UVW2 $\sim$6\%, UVW1 $\sim$5\%, U $\sim$2\%, B $\sim$11\% and V $\sim$4\%. These contributions suggest that the emission lines may affect the band flux. Considering stellar contribution along with the emission lines, the net contribution in each band are estimated to be $\sim$13.3\%, $\sim$10.5\%, $\sim$7.0\%, $\sim$16.5\%~and $\sim$9.8\%~ in UVW2, UVW1, U, B and V bands, respectively. We obtained the count rates and corresponding flux densities in each of the OM bands for all the observations listed in Table~\ref{obs_log}. We, therefore, subtracted above derived fractions from the observed flux densities to get the flux densities free from the contributions due to stellar and emission lines, in each band. Further, we dereddened the flux densities using Galactic reddening factor 0.0112 \citep{2011ApJ...737..103S} and intrinsic reddening factor 0.260 as estimated by \citet{2010ApJS..187...64G}. Finally, we created light curves for flux densities of each band as shown Fig.~\ref{uvopt}~(a).

\subsection{Temporal analysis}

\subsubsection{Optical, UV and X-ray light curves}
   The optical/UV emission in Fig.~\ref{uvopt}~(a) shows spectacular variability in all the bands. The UV light curves reveal a major variability event which appears to have lasted for about a  year. Starting from September 2002, the UV emission increased continuously, peaked in June 2003, and then declined.  The optical V and B band emissions appear to be delayed in the rising part, there being no sign of flux increase from the first to second observation. However, the V and B band flux increased starting from the second observation and peaked in the fourth observation as in case of UV bands.  We noticed that the variability profile of the B and V bands appear to be narrower than the UV band profiles. In particular, the B band variability profile is clearly narrower than the UVW2 band profile. Also the start-to-peak variability amplitude $F_{var}$ as determined from Equation ~\ref{eq:frac_var} is the highest for the UVW2 ($22.3\%$) band and decreases with the effective wavelength of other bands.

   We calculated background--corrected, net source count rates in the $0.3-10\kev$ (full), $2-10\kev$ (hard) and $0.3-2\kev$ (soft) bands from the cleaned EPIC--pn data obtained from each of the observations and created the X-ray light curves in the three bands as shown in upper three panels of Fig.~\ref{uvopt}~(b). Interestingly, the rapidly increasing X-ray fluxes peak in 2003 March, not in 2003 June as the optical/UV emission does, and then decrease slowly with a sharp jump upwards in the end. The X-ray~count rates vary rapidly from low to high (in 2002 September to 2003 March) by $242.7\%$ in the $0.3-10\kev$ band, $71.4\%$ in the $2-10\kev$ band and $324\%$ in the $0.3-2\kev$ band. Following the maximum, reduction in the count rates was noticed with a variation of $25.2\%, 15.2\%$ and $27.1\%$ in the full, hard and soft bands, respectively. The $0.3-2\kev$ count rates show variations higher than those of the $2-10\kev$ band. In the fourth panel of Fig.~\ref{uvopt}~(b), the hardness ratio is shown between the $2-10\kev$ and $0.3-2\kev$ bands. Clearly, the X-ray spectrum of 1H~0419--577 became softer with time. The X-ray spectrum appears to be softer at higher flux as shown in Fig.~\ref{xflux_comp}. Both the optical/UV emission and X-ray emission seem to vary in a similar way and appear to be correlated to each other.

\begin{figure*}
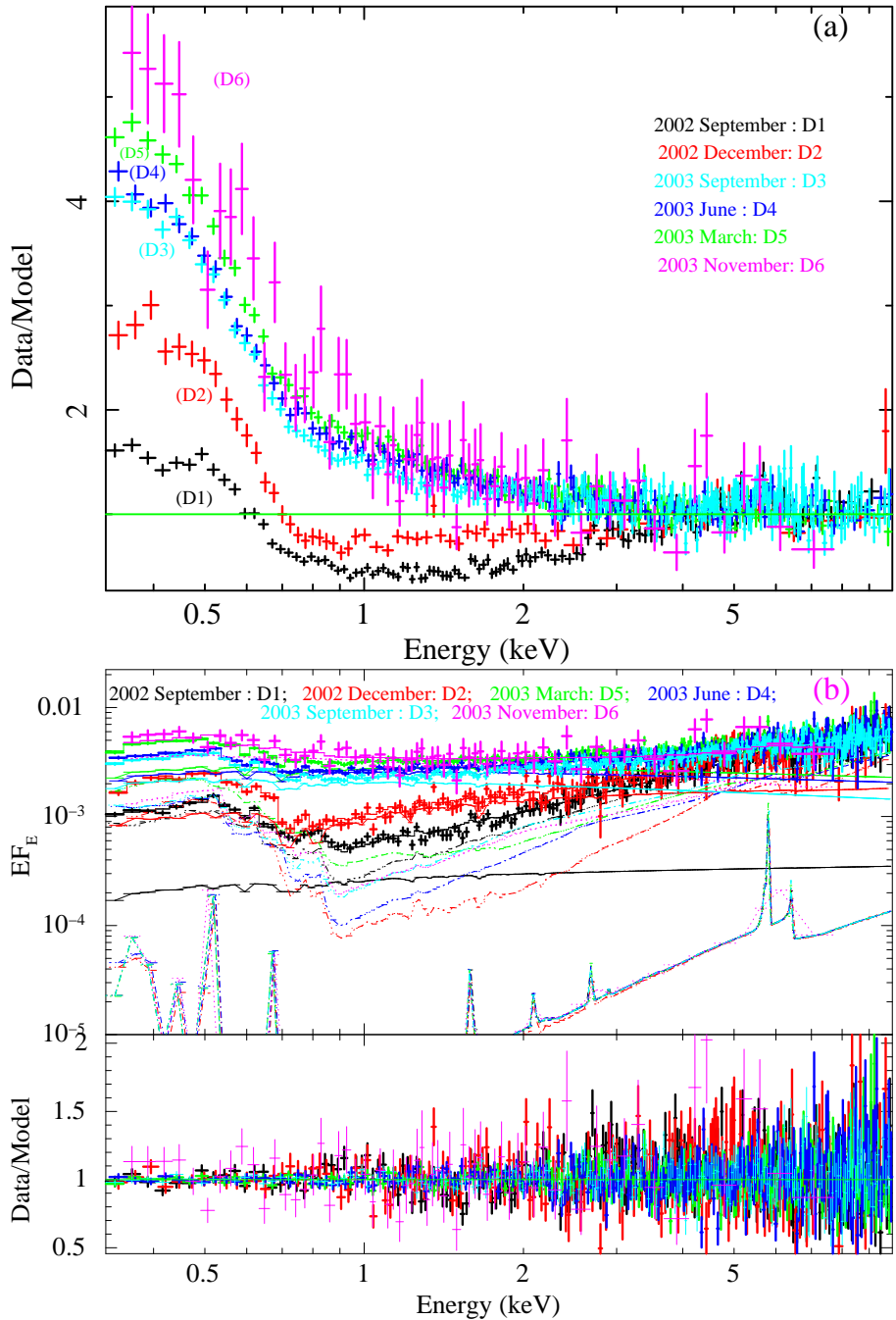

\includegraphics[scale=0.5,angle=-90.0]{F5a.eps}
  \includegraphics[scale=0.5,angle=-90.00]{F5b.eps}
\caption{ (a) The data to model ratio for observed data and fitted absorbed power law model. The absorbed power law is fitted with varying only its normalization in the $4-5$ keV to $7-10$ keV bands, and then added the $0.3-4$ keV and $5-7$ keV bands to see various features such as soft excess below 2 keV and possible broad iron line near 6 keV. (b) The best--fit model ({\tt xstar(relxill+nthcomp+xillver)}), model-components, data in upper panel and residuals (data to model ratio) in lower panel are shown. The dash-dot curves: distant reflection component (lower side of the upper panel); the solid lines: power law component and the dash-dot-dot-dot curves: blurred reflection component.} 
  \label{xmm_all}	
\end{figure*}

To compare the variability in different light curves, we
 made use of the fractional variability amplitude $F_{var}$
statistic \citep{2002ApJ...568..610E, 2003MNRAS.345.1271V} which is a
measure of the intrinsic variability corrected for measurement
uncertainty. The AGN variability being stochastic in nature, $F_{var}$ is not
useful for the light curves over non-simultaneous periods. However, for
simultaneous light curves in different energy bands $F_{var}$ provides a good
estimation of the variability characteristic. Here $F_{var}$ is given
by

\begin{equation}\label{eq:frac_var}
  F_{var} = {\frac{\sqrt{S^2 - \bar{\sigma}^2_{err}}}{\bar{x}}}
\end{equation}
where, S is the variance of the light curve, $\bar{\sigma}^2_{err}$ is
the mean square of the errors and $\bar{x}$ is the mean of the light
curve. The term $S^2 - \bar{\sigma}^2_{err}$ is also called as
\emph{excess variance} \citep{2002ApJ...568..610E} and is devoid of
measurement uncertainties.  The errors on $F_{var}$ are calculated
using equation B2 in the appendix given by \cite{2003MNRAS.345.1271V}. $F_{var}$ for optical/UV fluxes is in the range $3.6-7.3\%$. Similarly, $F_{var}$ is found to be $\sim 45\%$ and $\sim20\%$ for power-law flux in the 2-10 keV and soft X-ray excess flux in the 0.3-2 keV, respectively, obtained from spectral modeling described below. The rms spectrum from X-ray to UV/optical bands is shown in Fig.~\ref{fvar}.

\begin{figure}
\centering
\includegraphics[height=12cm,width=8.5cm]{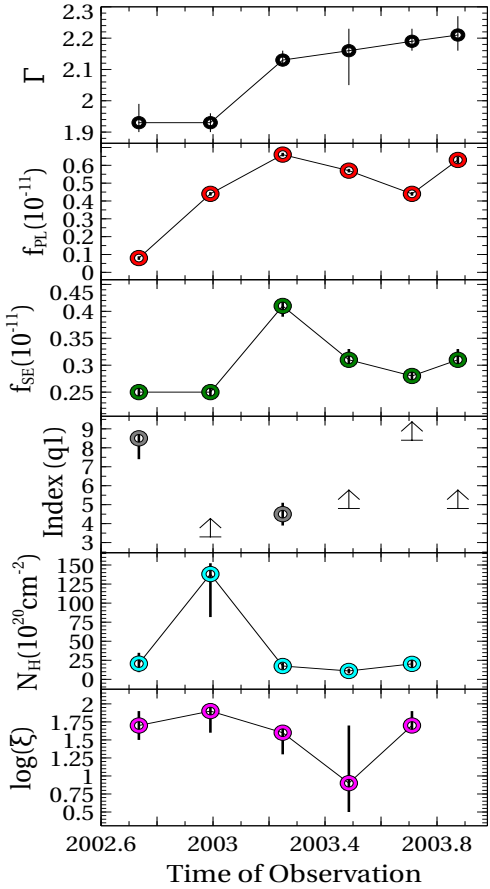}
\caption{ The light curves of six X-ray spectral parameters from top to bottom-- photon index , $2-10$ keV power-law flux, soft excess due to blurred reflection in the $0.3-2$ keV band, inner emissivity index $q1$, column density and ionization parameter of the warm absorber. The flux of power law and soft excess components are measured in units of 10$^{-11}$\funit~while ionization parameter has units of erg s$^{-1}$ cm$^{-1}$. The error-bars on power-law flux and soft excess flux are quoted at one $\sigma$ level while the error-bars on photon index is shown at 90\%~confidence level. }
\label{xflux_comp}
\end{figure}

\subsubsection{Spectral indices in UV/optical and X-ray bands}

To characterize the broadband spectral energy distribution (SED) in UV/Optical and X-ray bands, we adopted a model-independent way based on UV/optical spectral index $\alpha_{\rm{UV}}$ and the UV-to-X-ray spectral index $\alpha_{\rm{OX}}$. $\alpha_{\rm{OX}}$ can be estimated by using the following equation, 

\begin{equation}\label{alphaox}
\alpha_{OX}=\frac{log(\frac{L_{\nu}(2~keV)}{L_{\nu}(2500\AA)})}{log(\frac{{\nu}(2~keV)}{{\nu}_(2500\AA)})}=-0.3838~log(\frac{L_{\nu}(2500\AA)}{L_{\nu}(2~keV)})
\end{equation}
where $L_{\nu}$(2500\AA) and $L_{\nu}$(2~keV) are the monochromatic luminosity densities at 2500\AA~ and at 2 keV, respectively. We fitted the $f_{\lambda}=c\times\lambda^{-\beta}$ (where $c$ and $\beta$ are the normalization factor and slope, respectively) to the observed flux densities of various bands for each epoch. Following this procedure, we estimated the flux density at 2500 \AA~ for each epoch. Such modeling can also provide the UV spectral index $\alpha_{\rm UV}=2-\beta$ which can give the spectral variation of the AGN over all epochs. While fitting power law to the flux densities, we could not constrain the normalization parameter $c$ and hence we were unable to estimate the errors on the flux densities at 2500 \AA. Such normalization caused no error estimation while determing $\alpha_{\rm OX}$. Further, we also derived the flux density at 2 keV using the $0.3-10$ keV continuum flux from the broadband X-ray modeling. Flux density $F_{\nu}$ at a given frequency $\nu$ can be estimated by using the relation $F_{\nu}=k\nu^{-\alpha}$, $k$ and $\alpha$ being some constant and spectral index, respectively. Integrating above relation from $\nu_1$ to $\nu_2$, the flux density $F_{\nu}$ can be expressed as, 
\begin{equation}\label{fluxdensity}
F_{\nu}=\frac{(1-\alpha)~F~\nu^{-\alpha}}{\nu_{2}^{1-\alpha}-\nu_{1}^{1-\alpha}}
\end{equation}
where $F$ is the integrated flux between the frequencies from $\nu_1$ to $\nu_2$. Using estimated flux densities at 2 keV and 2500 \AA, we derived $\alpha_{\rm{OX}}$ for each epoch. The variation of $\alpha_{\rm{OX}}$ and $\alpha_{\rm{UV}}$ are shown in Fig.~\ref{oxuv}. 

\begin{table}
  \centering 
  \caption{Derived parameters and counts for AGN and host galaxy components. 
    \label{psfDecomp} }
  \begin{tabular}{lcccc}

    \hline
    \hline 
    & \multicolumn{2}{c}{ Moffat function} & \multicolumn{2}{c}{ Sersic function} \\
    Band                      &  White dwarf           &  AGN              & host galaxy             & host galaxy\\
                              & ($a,b$)                & counts        & counts     & contribution (\%)   \\

    \hline
    $ V$                      & (1.21,~2.24)     & 11025    &677.28   & 5.8   \tabularnewline
    $B$                       & (1.11,~2.03)     & 23620.6     &1364.3  & 5.5   \tabularnewline
    $U $                      & (1.76,~2.54)     & 20290.8    &1064.9  & 5.0  \tabularnewline
    $ UVW1$                   & (2.1,~2.65)     & 10229.1     & 589.5 & 5.5   \tabularnewline
    $ UVW2$                   & (3.18,~4.07)     & 993.5     & 78.9 &  7.3  \tabularnewline
     \hline
  \end{tabular} \\
 \end{table}

\begin{figure*}
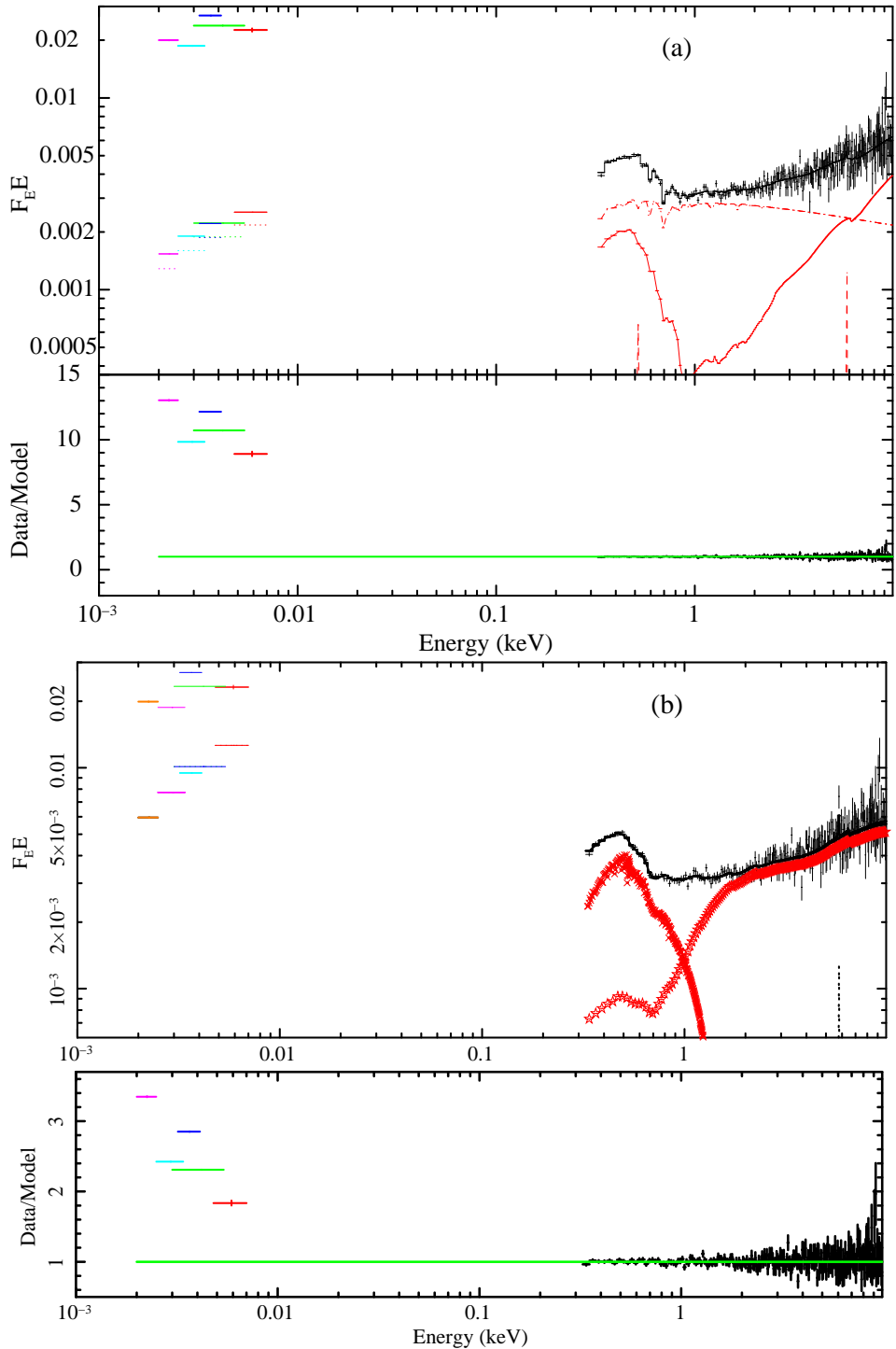

\centering
\includegraphics[scale=0.53,angle=-90.00]{F7a.eps}
\includegraphics[scale=0.54,angle=-90]{F7b.eps}
\includegraphics[height=13cm,width=4cm,angle=-90]{F7c.eps}
\caption{ The left side markers below 0.01 keV represent the UV/optical data and the right side above 0.01 keV black data points corresponds to the X-ray data in the top panels of each plot.  The best--fit of the X-ray band and the extrapolation down to the UV/optical band using the model (a) Top panel (reflection model with disc component): {\tt zredden$\times$redden$\times$xstar(relxill+nthcomp+xillver+diskpn)}, bottom panel: residuals in data to model ratio. The dash-dot line: power law component and the solid red curve: blurred reflection component  (b) Top panel (Intrinsic disc model): {\tt zredden$\times$redden$\times$zxipcf$\times$zxipcf(xillver+optxagnf)}, bottom panel: residuals in data to model ratio. The residuals in the UV/optical bands exhibit a clear excess. The cross red marker: soft Comptonization component and the star red marker represents the power law component. The dash-dot curves: distant reflection component (lower side in the upper panel of both plots) and the black solid lines: net resultant model.} 
\label{optxagnf}
\end{figure*}

\subsection{Spectral modeling}
We used {\tt XSPEC v12.9} \citep{1996ASPC..101...17A} to analyze the X-ray spectral data and used the $\chi^2$ minimization technique to find the best--fit model parameters for all the observations. Below we quote the errors on the best-fit parameters at $90\%$ ($\Delta \chi^{2}=2.71 \sigma$) confidence level unless otherwise specified.

We used the EPIC--pn spectral data  due to the high signal to noise ratio compared to EPIC--MOS data. In order to find possible spectral variations, we first compared the EPIC-pn spectral data sets from all the six observations. We fitted an absorbed power law ({\tt tbabs$\times$power~law}) model jointly to the six data sets in the $4-5$ keV and $7-10$ keV bands. To account for the Galactic absorption, we fixed the column density at $N_{\rm H}= 1.83\times10^{20}~\rm cm^{-2}$ \citep{dickey1990} in the {\tt tbabs} model which includes the latest photo-absorption cross-section and abundances \citep{2000ApJ...542..914W}. In the above energy bands, we varied the power-law normalization independently for each data set while we kept the photon index the same. The fit resulted in $\chi^{2}/dof= 285.6/291$, where dof stands for degrees of freedom, the best--fit photon index being $1.5\pm0.1$. We then included the $0.3-4$ keV and $5-7$ keV bands for further analysis in the $0.3-10$ keV band. We have shown the data to model ratio in Fig.~\ref{xmm_all}~(a), which clearly reveals spectral shape variability. The lowest state in 2002 September reveals a curvature down to 1\kev, a weak absorption feature near 1\kev,~possibly an unresolved transition array (UTA) (e.g, \citet{2001A&A...365L.168S}) feature of low ionized iron and a rising soft excess below 1\kev. The soft excess features seem stronger in 2002 December ( see Fig.~\ref{xmm_all}~(a)). Moreover, the soft excess appears stronger in the rest of the observations at low energies.

To characterize the broadband X-ray emission for all data sets, we begin with simple absorbed power law model in the $2-10$  keV band using EPIC-pn data set of 2002 September. The fit resulted in $\chi^{2}/dof= 217.4/159$. The best--fit photon index we found was $\Gamma=1.00\pm0.04$. The flat power-law index ($\Gamma\sim1$) in the fit is difficult to understand, though this may be possibly due to extreme smearing either by reflection (e.g., \citealt {fabian2005}) or absorption (e.g., \citealt{pounds2004a, pounds2004b}). In some studies, it has been suggested that photon starvation may be responsible for the flat photon index (e.g., \citealt{2002MNRAS.330L...1P}). However, the measurement of reverberation delays in a number of AGN suggests that blurred reflection is more likely (see e.g, \citealt{2013MNRAS.431.2441D,2011MNRAS.418.2642Z}). The blurred reflection model also describes the broadband X-ray spectra of 1H~0419--577 (see e.g., \citealt{ 2013MNRAS.435.1287P,2010MNRAS.408..601W}). We therefore used the relativistic reflection model {\tt relxill}  which is the combined form of the reflection model {\tt xillver} \citep{ 2011ApJ...731..131G, 2013ApJ...768..146G} and relativistic line model {\tt relline} \citep{2010MNRAS.409.1534D, 2013MNRAS.430.1694D}. This model calculates the reflected emission at each angle at each radius of the accretion disc \citep{2014ApJ...782...76G}. This model consists of the built-in power law component to illuminate the disc. The detailed parameters of {\tt relxill} and its different applications are described in brief on a webpage document\footnote{http://www.sternwarte.uni-erlangen.de/~dauser/research/relxill/relxill\_doc.pdf}.

\begin{table*}
\caption{The best--fit parameters of reflection model (xstar (XS.)$\times$(relxill (REL.)+nthcomp (NTH.)+xillver (XIL.)))
fitted to the 0.3-10 keV band. Due to the short exposure of 2003 November observation, we fixed the blurred reflection components to the best--fit parameters obtained for 2003 September. The flux is measured in units of $10^{-11}$\funit. ** represents that the values are common to all data sets.  ``T4" stand for those parameters tied with 2003 June (4th observation) data sets. `t' is used for the tied parameter for a data set. }
\label{fit_rel} 
\begin{tabular}{lcccccccccc}
\hline 
\hline 
\small 

Model  &Model component        &2002 September   &2002 December       &2003 March            &2003 June          &2003 September          &2003 November  \\
&&Low flux state& \multicolumn{4}{c}{Intermediate flux state}& High flux state  \\ 							
\hline
Gal. abs.&$N_{\rm H}$ ($10^{20}\rm cm^{-2}$)                                      &\multicolumn{6}{c}{1.83~(f)$^{**}$}  \\ 
NTH. &$\Gamma$    &$1.93_{-0.03}^{+0.06}$      &$1.93\pm0.03$         &$2.13_{-0.02}^{+0.03}$       &$2.16_{-0.11}^{+0.07}$             &$2.19_{-0.03}^{+0.04}$         &$2.21_{-0.05}^{+0.06}$           \\ 
&f$_{E}$~~($2-10\kev$) &$0.08\pm0.01$          &$0.44\pm0.01$         &$0.66\pm0.01$       &$0.57\pm0.01$             &$0.44\pm0.01$   &$0.63\pm0.02$ \\

XS.  &$N_{\rm H}$ ($10^{20}\rm cm^{-2}$)  
                   &$20.6_{-11.2}^{+14.2}$  &$138.1_{-56.4}^{+35.1}$   &$17.5\pm4.7$          &$11.0_{-3.3}^{+3.1}$ &$20.3_{-7.2}^{+6.0}$        &$0.1$~(f) \\
       &\logxi        &$1.7\pm0.2$             &$1.9_{-0.3}^{+0.1}$       &$1.6_{-0.3}^{+0.1}$   &$0.9_{-0.4}^{+0.8}$  &$1.7_{-0.1}^{+0.2}$   &$4$~(f) \\%
REL.   &$q1$           &$8.5_{-1.1}^{+0.3}$    &$>3.3$                &$4.5\pm0.6$             &$>4.8$  &$>8.4$                   &T4   \\
       &$q2$                                        &\multicolumn{6}{c}{3~(f)$^{**}$}   \\
 &$\Gamma$    &$1.93_{-0.03}^{+0.06}$ (t)      &$1.93\pm0.03$  (t)      &$2.13_{-0.02}^{+0.03}$ (t)       &$2.16_{-0.11}^{+0.07}$ (t)            &$2.19_{-0.03}^{+0.04}$ (t)         &$2.21_{-0.05}^{+0.06}$ (t) \\
&$\xi$ (\xiunit)      &$30.9_{-4.1}^{+6.5}$    &$23.7_{-6.7}^{+7.9}$  &$40.0_{-11.1}^{+5.8}$     &$10.7_{-2.9}^{+1.7}$ &$11.2_{-2.1}^{+2.6}$    &T4\\

   &$R_{in}$ ($r_{g}$)    &$<1.3$              &$3.3_{-0.9}^{+0.6}$    &$<1.9$                   &$1.7\pm0.3$           &$1.5\pm0.1$           &T4 \\
   &$R_{br}$ ($r_{g}$)    &$3.2\pm0.5$         &$>3.8$                 &$>3.1$                   &$3.3_{-0.4}^{+3.0}$  &$3.1_{-0.2}^{+0.1}$            &T4    \\
         &$a$                                  &\multicolumn{6}{c}{$>0.99$$^{**}$}   \\
         &$R_{frac}$                           &\multicolumn{6}{c}{$-1$~(f)$^{**}$}   \\
         &$i$ (degree)                         &\multicolumn{6}{c}{$30$~(f)$^{**}$}   \\
&f$_{E}$~~($0.3-2\kev$)   &$0.25\pm0.01$       &$0.25\pm0.01$          &$0.41_{-0.02}^{+0.01}$    &$0.31_{-0.02}^{+0.01}$        &$0.28\pm0.01$         &T4  \\
XIL.
&$\Gamma$                                          &\multicolumn{6}{c}{2.2~(f)$^{**}$}         \\
     &$\xi$ (\xiunit)                              &\multicolumn{6}{c}{$1$~(f)$^{**}$}   \\
     &$i$ (degree)                         &\multicolumn{6}{c}{$30$~(f)$^{**}$}   \\
     &Norm. ($10^{-6}$)                            &\multicolumn{6}{c}{$4.7_{-3.1}^{+3.5}$$^{**}$}   \\

Stat. &                                            &\multicolumn{6}{c}{$\chi^{2}/dof=1201.8/1107$}                   \\

\hline  
\end{tabular}
\end{table*}

  The applied form of {\tt relxill} in our analysis assumes that the X-ray source illuminates the accretion disc. The illumination is described as a broken emissivity law which has the form $\epsilon \propto r^{-q_{in}}$ between $r_{in}$ and $r_{br}$, and $\epsilon \propto r^{-q_{out}}$ between $r_{br}$ and $r_{out}$; where $r$ is the distance between accretion disc and X-ray source, $q1$ and $q2$ are inner and outer emissivity indices and $r_{in}$ , $r_{br}$ and $r_{out}$ are the inner, break and outer radii of accretion disc, respectively. The other parameters are spin of the central black hole $a$, inclination angle $i$, iron abundance $A_{Fe}$ relative to solar abundance, illuminating power-law index $\Gamma$, high energy cutoff $E_{cut}$, ionization parameter ($\xi=L/nr^{2}$, where $L$ is X-ray luminosity of source, $n$ is the hydrogen number density of disc material and $r$ is the distance of the X-ray source from the accretion disc) and reflected fraction of distant emission denoted by $R$.  The negative value of $R$ is used to get only reflected flux and then separate power law component is required to illuminate the disc. We added {\tt relxill} model in the $0.3-10$ keV absorbed power law fit (e.g, {\tt tbabs$\times$(relxill+power~law})).  We tied the {\tt relxill} photon index to power-law photon index $\Gamma$ and the  $R$ parameter for relative reflection fraction was fixed to $-1$. We also fixed the iron abundance to 1, the inclination to 30 degree, high energy cutoff to 500 keV and outer radius to $400r_g$. The fit with {\tt power~law+relxill} model resulted into $\chi^{2}/dof= 364.5/216$. 

         In order to account for reflection from distant optically thick matter such as the putative torus, we added {\tt xillver}. The model parameters are iron abundance relative to solar abundance, inclination, high energy cutoff, ionization parameter and its photon index for X-ray illumination. We fixed its photon index at 2.2, the inclination to 30 degree, high energy cutoff to 500\kev, iron abundance to 1, ionization parameter to $1~\rm ergs~s^{-1}~cm^{-1}$~and we varied only its normalization. The fit significantly improved to $\chi^{2}/dof= 274.5/215$. We further varied its ionization parameter which improved the fit by $\Delta \chi^{2}=-13.1$ with additional one free parameter.  

          To be more realistic, we replaced the phenomenological power law model by {\tt nthcomp} \citep{1996MNRAS.283..193Z, 1999MNRAS.309..561Z} which can correctly predict the low energy roll over where Galactic absorption can modify the spectrum. We fixed seed photon temperature at $4$ \ev~as obtained for maximally spin black hole with mass $3.8\times10^{8}~M_{\odot}$ for this AGN accreting with Eddington rate and electron temperature associated with X-ray corona to 200 \kev. The fit resulted in $\chi^{2}/dof= 259.7/214$. The residuals show a dip due to the OVII edge and a weak smooth UTA feature below 1\kev. \citet{pounds2004a, pounds2004b} confirmed the presence of the OVII complex edge using reflection grating spectrometer (RGS) data for this object. To model these warm absorber features, we created {\tt xstar} model grids assuming the illuminating power-law photon index $\Gamma=1.8$, seed photon temperature $T=10^{4}$K and luminosity $L=10^{44}$\ergsec. We, then, added the {\tt xstar} model to fit the warm absorber features. We allowed the variation of its column density $N_{\rm H}$ and ionization parameter $\xi$. The fit improved by more $\Delta \chi^{2}=-11.1$ with two extra free parameters. We did not find any significant residuals in the final fit. Thus the fit resulted in $\chi^{2}/dof= 247.8/211$. 

        The distant reflection {\tt xillver} parameters are expected not to vary significantly between various observations. In order to carry out the spectral variability of various spectral components such as soft excess and power-law emission, joint analysis is the best tool to fit all the data sets together. We, therefore, tied {\tt xillver} parameters of all data sets to that of 2002 September observation. We fixed the parameters of {\tt xillver} at $\xi=1$~\xiunit,~$A_{Fe}=1$~and $\Gamma=2.2$ and allowed its normalization to vary. The exposure of 2003 November observation is short and consists of usable data of only about {\color{blue} 0.3 ks}. To get power law component for this data set, we also tied the {\tt relxill} parameters, which is considered to provide less variable reflected emission compared to power-law emission (see e.g., \citealt{fabian2005}), to the {\tt relxill} parameters of similar flux state which is from 2003 June. We linked its {\tt relxill} parameters such as $q1$, $\xi$, $R_{br}$, $R_{in}$ and its normalization to the same observation. 

\begin{figure}
\includegraphics[height=10.5cm,width=9.4cm]{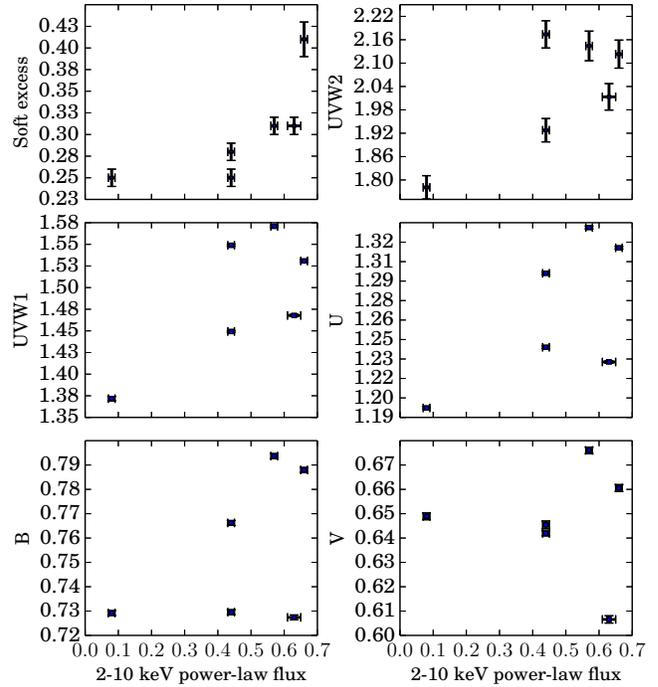}
\caption{ Flux-flux plots for correlation between the UV/optical flux densities and the 2-10 keV power-law flux. The UV/optical flux densities are shown as the function of power-law flux for soft excess to V bands from left to right. The unit of fluxes of soft excess and power-law components is $\rm 10^{-11}~erg~s^{-1}~cm^{-2}$ and the unit of flux densities in the UV/optical bands is measured as $10^{-14}$\efluxA. }
  \label{plox}	
\end{figure}

\begin{figure}
\includegraphics[scale=0.44]{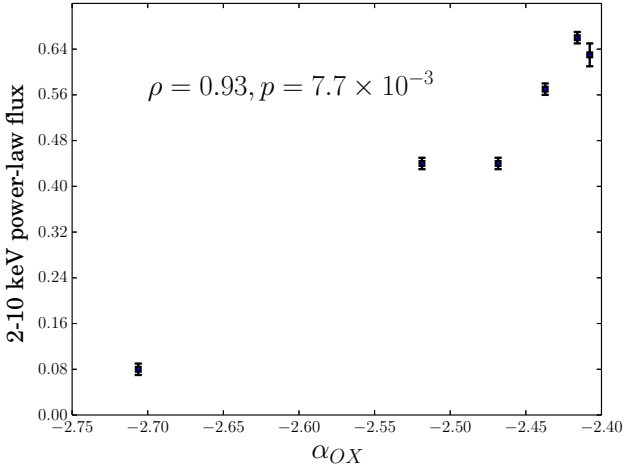}
\caption{The 2-10 keV power-law flux and the $\alpha_{\rm{OX}}$ are found to be strongly correlated.} 
  \label{plox1}	
\end{figure}

In the joint analysis, we used various model components to describe multiple data sets. Sometimes a model component is applicable for one data set but may not be required for other data set(s). For example, the UTA feature is visible in the 2002 September and 2002 December data sets, while it is hardly visible in other data sets. Initially, we found that the {\tt xstar} parameters for 2003 November observation are pegged to its hard and soft limits for ionization parameter (\logxi$=4$)~ and column density ($N_{\rm H}=10^{19}~\rm cm^{-2}$), respectively, suggesting not required for this data set. Therefore, we froze them to their limits. We allowed rest of the parameters of model components to vary. This final fit did not show any significant residuals and resulted in $\chi^{2}/dof= 1201.8/1107$. The best--fit data, model components and residuals are displayed in Fig.~\ref{xmm_all}~(b). The fit results are listed in Table~\ref{fit_rel}. We derived the {\tt nthcomp} flux in the $2-10$ keV band and soft excess flux in the $0.3-2$ keV band from {\tt relxill} model. The photon index and X-ray flux components are shown in Fig.~\ref{xflux_comp}.

\begin{figure*}\centering
\includegraphics[scale=0.45]{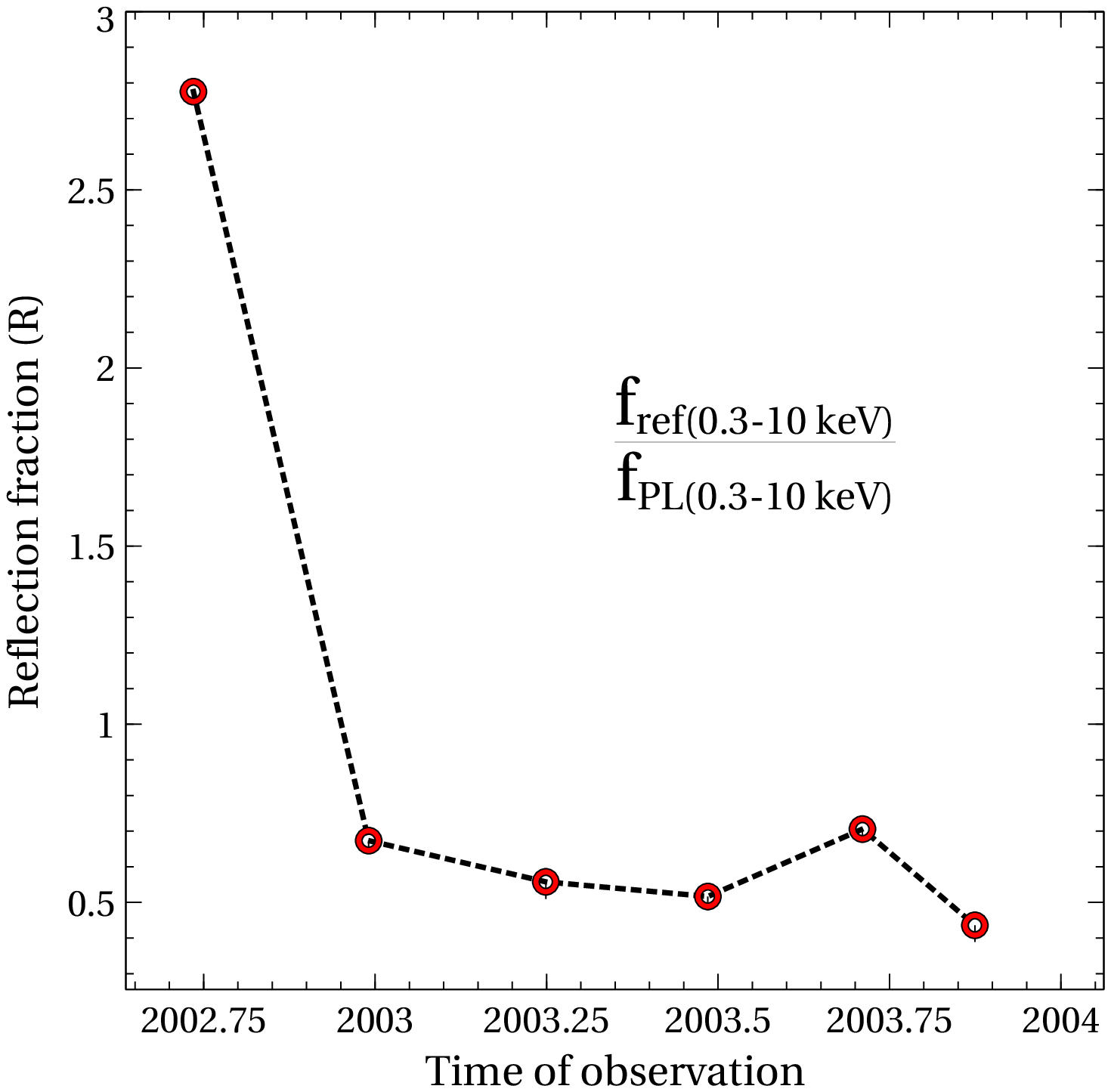}
\includegraphics[scale=0.47]{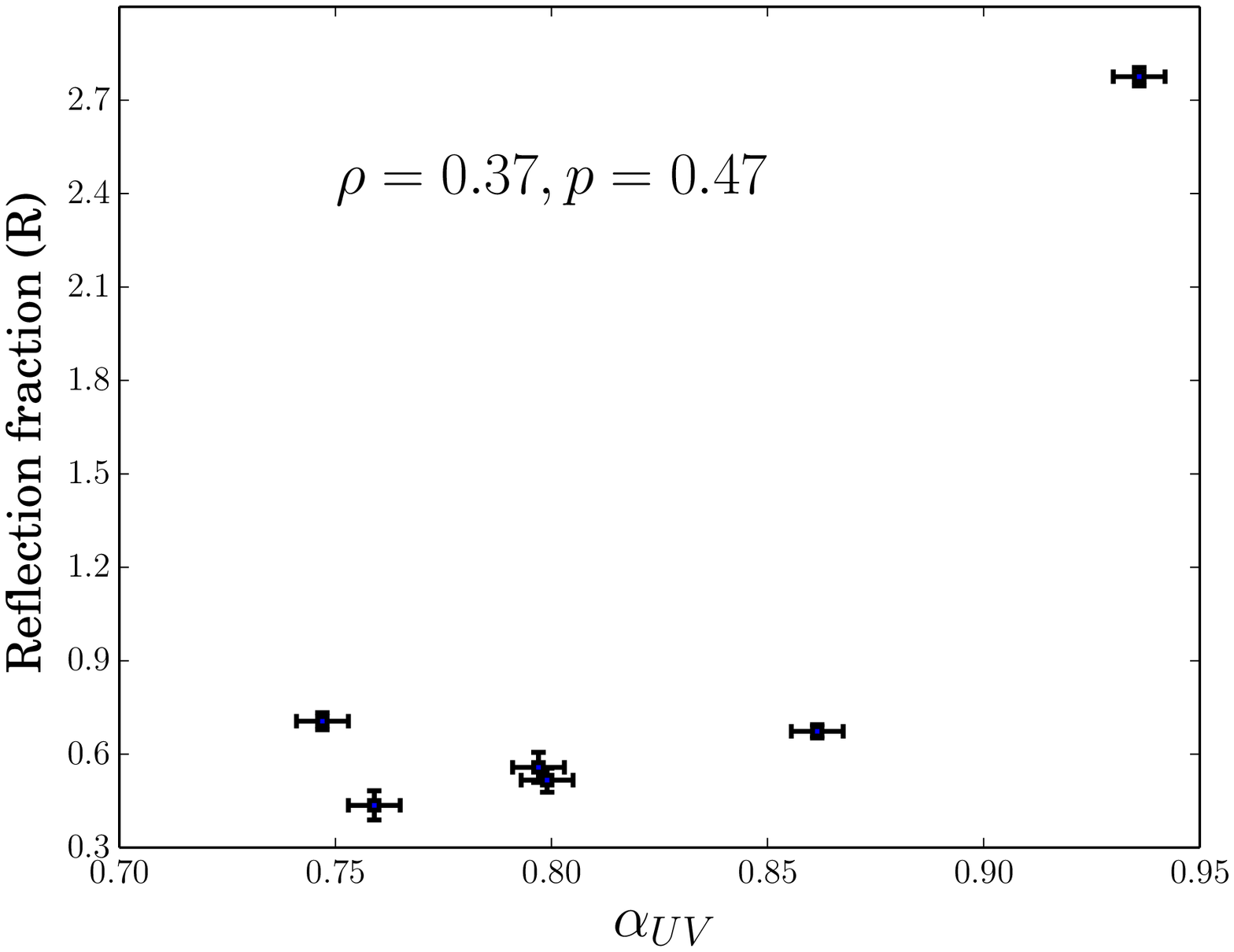}
\caption{ Change in the reflection fraction parameter with time (left panel) and the correlation between the reflection fraction parameter and $\alpha_{\rm{UV}}$ are shown (right panel).} 
  \label{R_UV}	
\end{figure*}

In order to investigate the connection between the UV/optical and  X-ray emission, we performed broadband optical/UV to X-ray spectral fitting. For this purpose, we used the spectral data obtained from the 2003 March observation, which has the largest net count (see Table~1). We used this count rate after subtracting the fraction of stellar contribution of host galaxy. We first fitted the X-ray band ($0.3-10\kev$) with the same model {\tt xstar$\times$(xillver+relxill+nthcomp)} as used in the joint spectral analysis. We fixed the parameters of the distant reflection to the best-fit values obtained from the analysis and varied other parameters as we did for the earlier analysis. The fit resulted in $\chi^{2}/dof= 201.8/220$. Further, we extrapolated the X-ray fit down to UV/optical range including the UVW2, UVW1, U, B and V bands. The UV/optical emission is expected from the disc and other constituents of the host galaxy. We therefore added {\tt diskpn} model to fit the UV/optical emission. This model consists of three parameters: temperature T$_{max}$, inner radius R$_{in}$, and its normalization $M_{BH}^{2}~cos (i)/D_{L}^{2}~\beta^{4}$, where $M_{BH}$ is mass in solar unit, $i$ is inclination, D$_{L}$ is the luminosity distance in kpc and $\beta$ is the colour temperature ratio. We calculated the {\tt diskpn} normalization to be $1.5\times10^{4}$ using the mass of the black hole $M_{BH}= 3.8\times10^{8}~M_{\odot}$ \citep{2005MNRAS.358.1405O}, $i=30$ degree \citep{fabian2005}, D$_{L}=500.5\times10^{3}~\rm kpc$ and $\beta=2.4$. We fixed the normalization at the estimated value and the inner radius at $6r_{g}$. We also fixed the inner temperature at $T_{max}=3\rm~eV$ as estimated for this source. We further dereddened the UV/optical bands by using {\tt redden} and {\tt zredden} being fixed at 0.0112 and 0.26. This results in a poor fit ($\chi^{2}/dof=3.26\times10^6/236$) and shows huge positive residuals suggesting an excess in the UV/optical emission compared to that expected from an accretion disc. Varying the normalization of {\tt diskpn} due to the uncertainty in mass and luminosity distance may help to describe the positive residuals (i.e., \citealt{2015MNRAS.450..857B}). The fit improved significantly to $\chi^{2}/dof=1.85\times10^6/235$ though there are still positive residuals present in the UV/optical bands. The broadband spectrum, the model and the residuals are shown in Fig.~\ref{optxagnf}~(a).

 We also tested the intrinsic disc Comptonized model, available as {\tt optxagnf} in XSPEC, for the accretion disc, soft excess and hard power law component  \citep{2012MNRAS.420.1848D}. This model describes the soft excess as the thermal Comptonization from an optically thick, cool plasma ($kT_e\sim0.3\kev$). The parameters for this model are the mass of black hole $M_{\rm BH}$, accretion rate relative to the Eddington rate $L/L_{Edd}$, black hole spin $a$, fraction of power that goes to hard X-rays above 2  keV $f_{pl}$, high energy power-law photon index $\Gamma$, the distance that separates the disc emission and coronal emission $r_{cor}$, the temperature of the cool plasma $kT_{e}$ and its optical depth $\tau$, outer radius of the disc and its normalization. We first fitted the X-ray band using {\tt optxagnf} model modified by the Galactic absorption. We fixed the mass at $3.8\times10^{8}~\rm M_{\odot}$ given by \citet{2005MNRAS.358.1405O}, spin parameter at $a=0.998$, an upper limit suggested from the analysis, outer radius at $10^5~r_{g}$ and its normalization to unity to get the correct flux from the disc. The naturally required distant reflection was fixed to its best-fit parameters as obtained from the joint analysis. We allowed rest of the parameters to vary. The fit resulted in $\chi^{2}/dof=435.6/225 $. 

The residuals showed the presence of mild absorption features in the soft X-ray band, which require further modification of the spectrum. The residuals required the {\tt zxipcf} model which is a {\tt xstar } photo-ionization code created with illuminating power-law photon index $\Gamma=2.2$ and the turbulence velocity 200 $\rm km~s^{-1}$ \citep{2008MNRAS.385L.108R}. This model has four parameters: redshift, covering fraction, ionization parameter and the column density. We fixed the redshift to the source redshift and varied column density, ionization parameter and covering factor of both the absorbers required for the fit. The final fit resulted in $\chi^{2}/dof=203.0/219$. We then fixed the accretion rate to the Eddington rate to get maximum disc emission and found that the fit did not change from its earlier best-fit and resulted in the same best-fit with $\chi^{2}/dof=203.0/220$. We then extrapolated the best-fit for the X-ray band down to UV/optical bands. After freezing the best--fit parameters of X-ray band, we added the {\tt redden} model to modify the full spectrum from our Galactic extinction. We fixed the reddening factor to $0.0112$ as obtained from \citet{2011ApJ...737..103S}. We also added the redshifted {\tt redden} model to modify the spectrum from intrinsic extinction within the host galaxy. The final fit resulted in a poor fit with $\chi^{2}/dof=1.13\times10^6/236$ due to the large excess in the UV/optical range. This is shown in Fig.~\ref{optxagnf}~(b).

\subsection{Correlation between various spectral components}

We studied the linear correlation among 2-10 keV power-law flux, flux of soft X-ray excess, UV/optical flux densities, $\alpha_{\rm{OX}}$, $\alpha_{\rm{UV}}$ and reflection fraction parameter. Since the data points are sparsely distributed, we used the Spearman rank-order correlation coefficient `$\rho$' to estimate the correlation between various spectral parameters. We determined correlation between the $2-10$ keV power-law flux and the UV/optical flux densities. We also derived $p$ values to calculate the significance of their correlation. The power-law flux and soft X-ray flux are strongly correlated with a correlation coefficient of 0.94 at 99.5\%~level. The correlation coefficients for the UVW2, UVW1 and U bands with the power-law flux are 0.38, 0.46 and 0.52, respectively, and their significance is at the modest level of $\sim60\%$. On the other hand, the correlation of B and V-bands with the 2-10 keV power-law flux are found to be very weak with correlation coefficient of 0.29 and 0.12, respectively (See of Fig.~\ref{plox}). The correlation between the 2-10 keV power-law flux and $\alpha_{\rm{OX}}$ is estimated to be 0.93 at a significance level of 99.2\% (see Fig.~\ref{plox1}). We also found poor correlation between the reflection fraction and the $\alpha_{\rm{UV}}$ (see bottom panel of Fig.~\ref{R_UV}) indicating complex origin of UV/optical emission in the AGN.

\begin{table*}\noindent \caption{Light-crossing, dynamical, thermal and viscous time scales for different UV/optical bandpass.} \label{times}

\begin{tabular}{ccccccc}
\hline
Optical/UV band     & Wavelength    &  Radius           & Light-crossing       & Dynamical        & Thermal      & Viscous  \tabularnewline
                    &  (\AA )       &  ($R_{S}$)        & Time scale (days)    &Time scale (days) &Time scale (days)       &Time scale ($10^{4}$ years)\tabularnewline

\hline
V                  & 5430          & 370.7             & 16.1           &438.9  &4389.0 &12.0\tabularnewline
B                 & 4390           & 279.2             & 12.1          & 286.9 &2868.8 & 7.9\tabularnewline
U                 & 3440           & 201.7             & 8.8          &176.2 & 1761.5 & 4.8\tabularnewline
UVW1              & 2910           & 161.4            & 7.0           &126.1 &1260.5 &3.5\tabularnewline
UVW2             & 2120            & 105.8            & 4.6           &66.9 &669.0 &1.8\tabularnewline
\hline
\end{tabular}
\end{table*}

\section{Discussion}
 
The UV/optical emission from 1H~0419--577 in different bands varied from minimum to maximum during 2002 September to 2003 June and then from maximum to a low value by 2003 November. Thus the \xmm{} monitoring observations covered a major variability event lasting for about a year. Though the variability pattern in different bands appear to be similar as the emission in all bands rises from a low value to maximum and again decreases from maximum to a low value, there are some differences. The fractional variability amplitude ($F_{var}=7.3-3.6$\%) in different bands decreases with the effective wavelengths of the optical/UV bands. Since shorter wavelength UV radiation arises from the inner accretion disc while the longer wavelength optical emission arises from relatively outer regions. This suggests that the $F_{var}$ increases with decreasing disc radius. In addition to the optical/UV emission, the strongest variability is observed for the $2-10\kev$ power law component arising from the innermost regions (see Fig.~\ref{fvar}). Wavelength-dependent variability characteristics may be intrinsic to the accretion disc/corona as predicted by the propagation fluctuation model. In such a case, X-ray emission must lag the UV/optical emission. However, the UV/optical variations appear to follow the variations in the X-ray emission. Also the variability timescale is much shorter than the viscous timescale. Hence, the UV/optical variability cannot be in the intrinsic emission from the accretion disc.

 The relationship between $\alpha_{\rm{OX}}$ and F$_{UV}$ at 2500 \AA~are expected to be anti-correlated for normal AGN (e.g., \citealt{2006MNRAS.368..479G}). However, our finding in 1H~0419--577 does not show such anti-correlation (see Fig.~\ref{oxuv}). This suggests that 1H0~419--577 is an X-ray weak AGN though the X-ray flux increases with time. X-ray weak nature infers the AGN to be reflection dominated where most of power-law emission is bent towards the disc (e.g., \citealt{2004MNRAS.349.1435M}). Comparing the derived $\alpha_{\rm{OX}}$~($<-2$ found in present work) to the $\alpha_{OX}\sim 1.5$ as found by \citet{2010ApJS..187...64G}, this AGN seems to be X-ray weak during 2002-2003. Similarly, $\alpha_{UV}\sim-0.65$ (as reported by Grupe et al. 2010), which is very low compared to what we found (see right panel of Fig.~\ref{oxuv}). This suggests that the source is also weaker in UV/optical bands during 2002-2003. Interestingly, the strong correlation between $\alpha_{\rm{OX}}$ and 2-10 keV power-law flux (see Fig.~\ref{plox1}) indicates that the UV/optical fluxes vary slowly compared to the X-ray flux and this is clearly seen in Fig.~\ref{fvar}. Such behavior hints that the X-ray and UV/optical emission are related to possibly a single radiation mechanism such as reprocessing phenomenon.

The possible delay of the optical/UV emission can, however, be caused by reprocessing of X-ray emission from the hot corona into the accretion disc. Thus, the observed optical/UV emission is most likely dominated by the reprocessed X-ray emission. This is indeed supported by our optical/UV to X-ray spectral modeling where we found that the optical/UV emission cannot be entirely described by the accretion disc component alone, and is in excess of that expected from the disc (see Fig.~\ref{optxagnf}). In the reprocessing scenario, the $2-10$ keV power-law flux obtained from the broadband spectroscopy is expected to correlate with each UV/optical emission  component. The observed moderate significance for  all the UV/optical bands (see Fig.~\ref{plox}) and, the reflection fraction parameter and $\alpha_{\rm{UV}}$ (see Fig.~\ref{R_UV}) may suggest that these bands are possibly either affected by additional components such as emission lines or due to the complex interplay between the accretion disc and the compact corona.

 Comparing the variability patterns of the optical and UV bands, it is obvious that the optical B and V bands appear to lag behind the UV bands -- UVW2, UVW1 and U in the rising part of the variability event (see Fig. 1). This suggests that the optical emission arises from relatively outer regions compared to the UV emission. Such time lags dependent on the wavelength of radiation are expected in the case of reprocessing of X-rays in the standard accretion discs with known temperature profile. Observations of wavelength-dependent time lags play a major role in probing the nature of accretion discs in AGN ( see e.g., \citealt{2014MNRAS.444.1469M,2016ApJ...821...56F, 2017MNRAS.466.1777P}). In the case of 1H~0419--577, it is not possible to determine the time-lag spectrum due to the sparse sampling of the data. Future multiwavelength monitoring observations such as those possible with the \astrosat{} mission will play an important role in determining the nature of accretion disc in 1H~0419--577.

 The expected time lag or lead between the X-ray emission and the UV/Optical emission is equivalent to the light-crossing time. The light-crossing time in a simple lamppost geometry is the time travel by the X-ray emission from a central compact corona to the region of the accretion disc peaking in the UV/Optical bandpass. Assuming the viscous heating, the effective wavelength $\lambda_{eff}$ of a bandpass can be converted into the temperature of a blackbody using the Wein's law. The temperature of geometrically thin and optically thick standard disc allows us to determine the radius of the disc corresponding to peak emission at the effective wavelength of a bandpass (e.g., \citealt{2007MNRAS.375.1479S}). Thus approximate light-crossing time at a given radius between the disc and the compact corona can be expressed as

\begin{equation}
\tau \approx 3.0\left(\frac{\lambda_{eff}}{3000\AA}\right)^{4/3}\left(\frac{\dot M}{\dot M_{Edd}}\right)^{1/3}\left(\frac{M_{BH}}{10^8}\right)^{2/3}~\rm days
\end{equation}
where $\frac{\dot M}{\dot M_{Edd}}$ is the ratio of accretion rate to the Eddington rate and black hole mass $M_{BH}$ is used in unit of $10^{8}~M_{\odot}$. We calculated the light-crossing time for each effective wavelength for respective bandpass and listed in Table~\ref{times}. We can derive other time scales associated with the accretion disc such as dynamical, thermal and viscous time scales (e.g., \citealt{2008NewAR..52..253K}). The shortest time scale is the dynamical time scale ($t_{dyn}$) corresponding to the Keplerian frequency.
\begin{equation}
t_{dyn}=\left(\frac{r^{3}}{GM_{BH}}\right)^{1/2} \approx 500\left(\frac{M_{BH}}{10^8~M_{\odot}}\right)\left(\frac{r}{r_g}\right)^{3/2}~\rm sec
\end{equation}
here $r_g$ is the gravitational radius. The thermal time scale ($t_{th}$) is related to the dynamical time scale ($t_{th}=\frac{t_{dyn}}{\alpha}$) and the viscous time scale can be defined as $t_{vis}=t_{th}\left(\frac{r}{h}\right)^2$, where $\alpha$ and $h$ are the viscosity parameter and height of the disc, respectively. We listed all the time scales for a simple comparison corresponding to the peak emission of a bandpass in the Table~\ref{times} assuming $\frac{\dot M}{\dot M_{Edd}}=1$, $M_{BH}=3.8\times10^8~M_{\odot}$~\citep{2005MNRAS.358.1405O}, $\alpha=0.1$, $\frac{h}{r}=0.01$ (e.g., \citealt{2013peag.book.....N}). The light-crossing time scale is the shortest compared to the dynamical, thermal and viscous timescales. 

The light-crossing time scale for 1H~0419--577 ranges from $\sim5$ to $\sim16$ days. The observed profile of the UV/optical emission seem to follow the X-ray emission possibly on about three month time scale. This duration is much longer compared to the estimated light-crossing time scale. The maximum exposure of a particular observation used in this work is $\sim15$ ks. This duration is much shorter than the expected light- crossing time scale. Therefore, the dominance of X-ray illumination over the accretion disc is difficult to predict. On the other hand, when the UV/optical photons act as the seed photons to give the X-ray emission by inverse Compton scattering, the UV/Optical emission and the X-ray emission must be connected on light-crossing time scale. In the Comptonization process, the UV/Optical emission lead the X-ray emission. This is contrary to the seen trends between the UV/optical band and the X-ray bands. \citet{2014A&A...563A..95D} presented the joint fitting in the UV/Optical and the X-ray bands using simultaneous the UV/Optical and the X-ray emission with $\sim$ one day long observation. They claimed that  the UV/Optical emission may give the X-ray emission by the Comptonization phenomena. Though this fit raises a concern how the UV/Optical emission can be related through $\sim$ a day long observation whose duration is much shorter than the expected light-crossing time.

 Connection between the UV and optical bands in the declining part of the variability event appears far more complex. The variability profiles in the UV bands are generally broader compared to those in the optical bands. The UVW2 band is clearly delayed compared to the optical B and V bands in the declining phase. This is contrary to the expectations from the simple disc reprocessing model where the optical emission arising from outer regions is expected to lag behind the UV emission arising from the inner regions. One possibility is that there is relatively stronger contribution of an additional delayed component such as the broad emission lines in the UVW2 band. Since the broad line region (BLR) is larger than the UV continuum emitting region as is known from the reverberation mapping of a number of AGN ( see e.g., \citealt{2011ApJ...735...80Z, 2005ApJ...629...61K,2004ApJ...613..682P, 2000ApJ...533..631K}), the contribution of broad emission lines in any band will cause additional delay in addition to the delay between the continuum in two different bands. 

Using the typical band-width of the OM filters, we found the net fraction of emission lines in $\rm UVW2\sim6\%, UVW1\sim5\%, U\sim2\%, B\sim11\%~ and~ V\sim4\%$. The net contribution in each band suggests that the emission lines flux may affect the band flux. Further, the emission line variability may also be reflected in the observed variable UV/optical emission if these emission lines in BLR arise likely as a result of photo-ionization of BLR by the continuum emission. 

The size of BLR ($R_{BLR}$) may help to understand the observed UV/optical variability. To estimate the size, we need monochromatic luminosity at 5100~\AA ($\lambda L_{\lambda}(5100\,{\mbox{\AA}}$) according to the following equation given by \citet{2000ApJ...533..631K}:
\begin{equation}\label{blr}
R_{BLR} = \left(32.9^{+2.0}_{-1.9}\right)\left(\frac{\lambda
L_{\lambda}(5100\,{\mbox{\AA}} )}{\rm 10^{44}\,erg\,s^{-1}}
\right)^{0.700\pm 0.033}\,{\rm lt\,days} 
\end{equation}

We derived the monochromatic luminosity using the best-fit linear model to the observed average flux densities in various bandpasses. The flux density from the best-fit linear model ( $(-4.2\pm0.7)\times10^{-4}\times\lambda$+$2.8\pm0.3$, where $\lambda$ is in $\AA$) is found to be $\sim 0.6\times10^{-14}$\efluxA~at $5100\AA$. The corresponding luminosity $\lambda L_{\lambda}(5100\,{\mbox{\AA}})$ is estimated to be $\sim9\times10^{44}$\ergsec. Thus using Equation \ref{blr}, we found a rough estimate of BLR size $\sim150$ days which is similar in order of magnitude to the observed delay between the X-ray and UV/optical emission. This may suggest that some of the variability in the UV/optical emission may be caused by BLR region. 

In addition to above, the UV light curves are much broader compared to optical light curves, and it seems that the UV emission is delayed compared to the optical emission. One would expect that the natural cause of these changes between UV and optical emission may be related to the variations in accretion flow, which are modulated at each radius towards the centre. Such variations, eventually, appear in highly variable X-ray emission which originates in the vicinity of the SMBH. In such a scenario, the X-ray emission is expected to lag the UV emission. However, the X-ray emission is narrower and is received earlier than the UV/optical emission. This suggests that the UV emission is partly contributed by the changes in accretion flow. 

 Alternatively, the observed UV/optical emission variability may be associated with the complex changes in the geometry of the X-ray corona, which is reflected in the soft excess emission. The soft excess flux changes drastically during 2003 March observation. This is difficult to understand unless there is no change in the coronal geometry. The UV/optical emission appear to delay compared to the X-ray emission and both the UV/optical emission show variations differently after the peak. The optical emission varies much faster than the UV emission. This may be because the X-ray corona expands initially while all emission components increase from minimum to maximum. Moreover, during the variations from peak to next low value, the X-ray corona shrinks, which can imprint further changes in illumination over the outer accretion disc, where the optical emission is expected to arise. In such a case, the X-ray emission is also expected to decrease which is not seen in the observed X-ray emission. The higher X-ray emission is possible when the X-ray corona is distributed to a greater extent vertically than radially.     

  Indeed, our study provides good insights in the complexities in X-ray to UV/optical emission variability. It helps to understand the possible scenario causing the variability. However, all observations are of short exposures and hence this object requires good signal to noise observations with proper monitoring in the X-ray to UV/optical bands. Such a programme is possible with the newly launched first Indian astronomical satellite {\it ASTROSAT}, a space facility to observe simultaneously over a broad range from optical to hard X-rays.

\section{Acknowledgment}
We sincerely thank the anonymous referee for his/her suggestions/comments which improved the manuscript significantly. This research has made use of archival data of \xmm{}  observatory, an ESA science mission directly funded by ESA Member States and NASA by the NASA Goddard Space Flight Center (GSFC). We also used the software and/or web tools obtained from the High Energy Astrophysics Science Archive Research Center (HEASARC) of the NASA/GSFC Astrophysics Science Division and of Smithsonian Astrophysical Observatory's High Energy Astrophysics Division.

\newcommand{\pasp}{PASP} \def\apj{ApJ} \def\mnras{MNRAS}
\def\aap{A\&A} \def\apjl{ApJ} \def\aj{aj} \def\physrep{PhR}
\def\pre{PhRvE} \def\apjs{ApJS} \def\pasa{PASA} \def\pasj{PASJ}
\def\nat{Nat} \def\ssr{SSRv} \def\aapr{AAPR} \def\araa{ARAA} \def\nar{NewAR}
\bibliographystyle{mn2e} 

\bibliography{refs}

\end{document}